\documentclass[aps,pre]{revtex4-1}
\usepackage{graphicx}
\usepackage{caption}
\usepackage{subcaption}
\usepackage{dcolumn}
\usepackage{bm}
\usepackage{amsmath}
\usepackage{amssymb}
\usepackage{epsfig}
\usepackage{verbatim}
\usepackage{pstricks,pst-grad,pst-plot}
\usepackage{natbib}
\usepackage{url}
\usepackage{makeidx}
\usepackage{wrapfig}
\usepackage{float}

\definecolor{lightyellow}{cmyk}{0,0,0.3,0}
\definecolor{lightblue}{cmyk}{0.1,0,0,0}

\begin{document}
\title{Effects of local and global network connectivity on synergistic epidemics}

\author{David Broder-Rodgers}
\affiliation{Selwyn College and Cavendish Laboratory,
 University of Cambridge, Cambridge, UK}
\email{db560@cam.ac.uk}

\author{Francisco~J.~P{\'e}rez-Reche}
\affiliation{Institute for Complex Systems and Mathematical Biology, SUPA, University of Aberdeen, Aberdeen, UK}
\email{fperez-reche@abdn.ac.uk}

\author{Sergei~N.~Taraskin}
\affiliation{St. Catharine's College and Department of Chemistry,
University of Cambridge, Cambridge, UK}
\email{snt1000@cam.ac.uk}

\begin{abstract}

Epidemics in networks can be affected by cooperation in transmission of infection and also connectivity between nodes.
An interplay between these two properties and their influence on epidemic spread are addressed in the paper.
A particular type of cooperative effects (called synergy effects) is considered, where the   transmission rate between a pair of nodes depends on the number of infected neighbours.
The connectivity effects are studied by constructing networks of different topology, starting with  lattices with only local connectivity  and then with networks which have both local and global connectivity obtained  by random bond-rewiring to nodes within certain distance.
The susceptible-infected-removed epidemics were found to exhibit several interesting effects: (i) for
epidemics with strong constructive synergy spreading in networks with high local connectivity, the bond rewiring has a negative role on epidemic spread, i.e. it reduces invasion probability;
(ii) in contrast, for epidemics with destructive or weak constructive synergy spreading on networks of arbitrary local connectivity, rewiring helps epidemics to spread;
(iii) and, finally, rewiring always enhances the spread of epidemics, independent of synergy, if the local connectivity is low.

\end{abstract}


\maketitle
\date{\today}

\section{Introduction}
 \label{sec:Introduction}

Dynamical processes on networks is a subject of broad interdisciplinary interest and intensive study~\cite{Newman_2010:book,Barrat_08:book}. In particular, network models provide a unique framework to describe a wide range of spreading processes including spread of infectious diseases, social phenomena or biological species~\cite{Barrat_08:book,Castellano_09:review,Fessel_PRL2012_FungalNetworks}. Such models assume a graph representation of systems with nodes (vertices) that can be in several states specific to the spreading process (e.g. a host infected by a pathogen or a patch occupied by some species). The state of nodes can change due to interactions with other nodes connected by links (edges or bonds) to a recipient. For instance, such interactions may represent transmission of infection, opinions, behaviours or ecological migrations.

The chances for a spreading phenomenon to affect a large number of nodes in a network, i.e. invade a network, depend on both the dynamics of interaction between nodes and the topology of the network~\cite{Liggett_85:book,Marro_99:book,Barrat_08:book,Durrett_PNAS2010,Diekmann_book2013,Pastor-Satorras_15:review}.
The simplest type of interaction is a pair-wise interaction when only two nodes are involved in transmission, e.g. an infected donor and a susceptible recipient.
The process of transmission can be characterised by two parameters, the rate of transmission (or probability of transmission in discrete-time description) and time of interaction (time of existence of contact/link between the donor and recipient) which can be a random variable.
If the transmission rate  is a fixed constant parameter then the transmission of infection is a homogeneous Poisson process, i.e. {\it simple transmission}.
However, in real situations the transmission can be a much more complex process.
In particular, both the life-time of nodes in different states and  the transmission rates can depend not only on characteristics of the donor-recipient pair but also on characteristics of other nodes.
In other words,  cooperative effects due to multiple-node interactions, which are called  synergistic effects below, can affect the values of parameters characterising transmission.
For example, the transmission rate can depend on the number of infected neighbours of a recipient.
This number can change with time throughout the course of an epidemic and thus the transmission rate can change with time.
These abrupt step-wise changes in transmission rate due to cooperative effects have significant and non-trivial effects on the spreading process.

The role of the network topology in the ability of epidemics to invade the network can be of crucial importance.
For example, assuming simple transmission, an invasion can be much easier in globally connected  networks such as complete or random graphs than in lattices where the nodes are locally connected to their nearest neighbours in space. However, it is not  clear \emph{a priori}   what the effects of cooperative or interfering synergistic phenomena in  transmission  would be on invasion in networks of different topology.
For example, if the local connectivity in lattices is reduced by the rewiring of some bonds which enhances  global connectivity, would this necessarily result in increased invasion ability for epidemics with synergistic effects?
This and other related questions are addressed in the paper.
Before going to the description of our model, we give a brief overview of existing models accounting for simple and complex transmission in spreading phenomena in networks of different topology.

Models assuming \emph{simple transmission} dynamics have provided good insight into some aspects of the interplay between the features of transmission and network topology. This is indeed the case for network models for epidemic spread, which often assume that the transmission of infection between a pair of donor-recipient nodes is independent of the rest of nodes connected to the pair \cite{Grassberger_83,Newman_2002:PRE}. A similar assumption was made in some models for spread of social phenomena~\cite{Zanette_2001:PRE}.
These models predict that invasions are facilitated in networks with small local clustering~\cite{Eguiluz-Klemm_2002:PRL,Petermann_2004_ClusteringEpidemics:PRE,Newman_2009_Clustering:PRL,Zanette_2001:PRE} and long links that can act as bridges for transmission of infection~\cite{Watts_Strogatz:1998,Newman_99:PRE,Moore_00,Zanette_2001:PRE}.
The size of invasions with simple transmission is therefore minimised in regular lattices (with relatively large clustering and no shortcuts) and maximised in random graphs (small clustering and many shortcuts). Small-world (SW) network topologies bridge the gap between lattices and random graphs by means of  either random bond  rewiring~\cite{Watts_Strogatz:1998,Watts:2003,Barrat_EPJB2000} or adding random shortcuts~\cite{Newman_99:PRE,Moore_00}.
The size and chance of simple invasions increase with the probability of either rewiring (i.e. by reducing local and increasing global connectivity) or adding shortcuts (increasing global connectivity)~\cite{Watts_Strogatz:1998,Newman_99:PRE,Moore_00,Sander_02,Santos2005,Khaleque_2013,Grassberger_2013:J_Stat_Phys}.

Many social and biological systems involve complex transmission dynamics which are often characterised by synergistic effects for donor-recipient pairs of nodes.
These effects are not captured by simple epidemiological models but they can significantly change the dynamics of spreading processes.
The knowledge and understanding of complex transmission dynamics on spreading and its interplay with the network topology is rather limited and is a topic of active research~\cite{Gross_2006:PRL,Centola_2007:PhysicaA,Centola_2010:Science,Montanari_2010:PNAS,Durrett_PNAS2010,Perez_Reche_2011:PRL,Ludlam_2011,Lu_2011:NewJPhys,Taraskin-PerezReche_PRE2013_Synergy,ZhengLuMing_2013_SocialReinforcement:PRE,Guo_2013:PRE,Gleeson_2013:PRX,Zhang-ChengLai_2014:Chaos}.

 Synergistic effects can be either constructive or interfering (destructive). Constructive synergistic effects from the neighbourhood of a recipient node were explicitly observed in experiments on the spread of behaviour~\cite{Centola_2010:Science} and fungal invasion~\cite{Ludlam_2011}. Social reinforcement was proposed as a key synergistic effect making invasions of social phenomena more likely and larger in clustered networks than in random graphs (i.e. opposite to the predictions obtained assuming simple transmission). This conclusion was supported by models involving social reinforcement from multiple neighbours~\cite{Centola_2007:PhysicaA,Montanari_2010:PNAS,Lu_2011:NewJPhys,ZhengLuMing_2013_SocialReinforcement:PRE}. The authors of Ref.~\cite{Lu_2011:NewJPhys} went a step further suggesting that these types of invasions are, in fact, optimal on SW networks rather than in fully clustered lattices. Interfering synergistic effects associated with, e.g. behavioural responses to epidemic spread~\cite{Zhang-ChengLai_2014:Chaos,Gross_2006:PRL,Guo_2013:PRE,Gleeson_2013:PRX} or competition for resources~\cite{Perez_Reche_2011:PRL}, can also play an important role in spreading dynamics.

Constructive and interfering synergistic effects are often described separately.
A recently developed model~\cite{Perez_Reche_2011:PRL} provides a flexible framework to study any degree of constructive or interfering synergy in any type of network. With this model, it was shown that synergy affects the size, duration and foraging strategy of spreaders~\cite{Perez_Reche_2011:PRL,Taraskin-PerezReche_PRE2013_Synergy} and can even result in explosive invasions~\cite{Gomez-Gardenes_Lotero_Taraskin_FJPR2015} of epidemics with and without node removal and for the Maki-Thompson model~\cite{Maki-Thompson_Book1973} describing social phenomena.
For regular lattice models, it was found that synergistic effects on invasion are enhanced by increasing local connectivity~\cite{Taraskin-PerezReche_PRE2013_Synergy}.

In this paper, we study the combined effect of local and long-range connectivity on synergistic spread. To this end, we use SW network models with rewiring which account for ubiquitous geographical constraints present in many social and biological systems~\cite{Barthelemy_2011:Phys_Rep}.
We study a synergistic SIR process on such networks.
The SIR model was originally formulated to investigate the spread of infection in populations where infected hosts either die or become permanently removed.
In this model, the nodes can be in three states: susceptible (S), infected (I) or removed (R).
For spread of social trends (e.g. opinion or rumour), similar states can be
used to distinguish between ignorant individuals (analogous to S), individuals that spread the trend (analogous to I) and individuals that stopped spreading (R).
We demonstrate that synergistic spread is strongly affected by the network topology. It is found that, in agreement with studies on social reinforcement~\cite{Centola_2010:Science,Centola_2007:PhysicaA,Montanari_2010:PNAS,Lu_2011:NewJPhys,ZhengLuMing_2013_SocialReinforcement:PRE}, systems with significant rewiring tend to be more resilient to invasion of epidemics with sufficiently constructive synergy. In contrast, interfering synergistic spread tends to be more invasive in rewired networks. We show, however, that these typical trends are very much affected by local connectivity. In particular, and in contrast to results in~\cite{Centola_2010:Science,Centola_2007:PhysicaA,Montanari_2010:PNAS,Lu_2011:NewJPhys,ZhengLuMing_2013_SocialReinforcement:PRE}, we show that rewiring systematically leads to larger invasions for weak local connectivity. We illustrate these and other effects with numerical simulations and analytical results for a simple model~\cite{Taraskin-PerezReche_PRE2013_Synergy} based on an approximate mapping SIR synergistic spread to uncorrelated dynamical percolation (such mapping is exact in the absence of synergy~\cite{Grassberger_83,Henkel_Hinrichsen_Book2009,kuulasmaa1982,Kuulasmaa_84}).

The structure of the paper is the following.
The model is introduced in Sec.~\ref{sec:Model} and results of its numerical analysis are given in Sec.~\ref{sec:Results}.
The analytical results are presented and compared with the results of numerical
simulations in Sec.~\ref{sec:Analytics}.
The conclusions are given in Sec.~\ref{sec:Conclusions}.
Some technical details are discussed in Apps.~\ref{app:AA},~\ref{app:A}, ~\ref{app:B} and~\ref{app:C}.

\section{Model}
\label{sec:Model}

\begin{figure}[ht]
\centering
\begin{subfigure}[b]{0.4\textwidth}
\centering
{\includegraphics[clip=true,width=\textwidth]{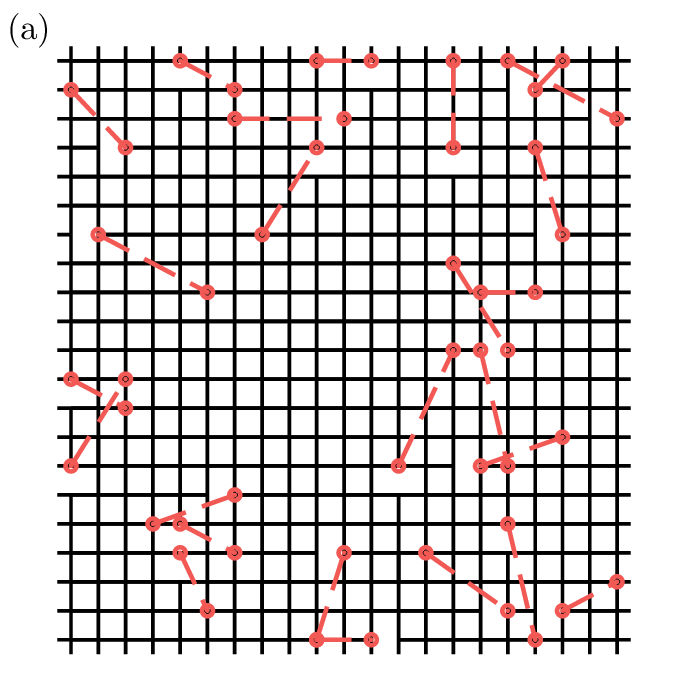}}
\end{subfigure}
\quad
\begin{subfigure}[b]{0.4\textwidth}
\centering
{\includegraphics[clip=true,width=\textwidth]{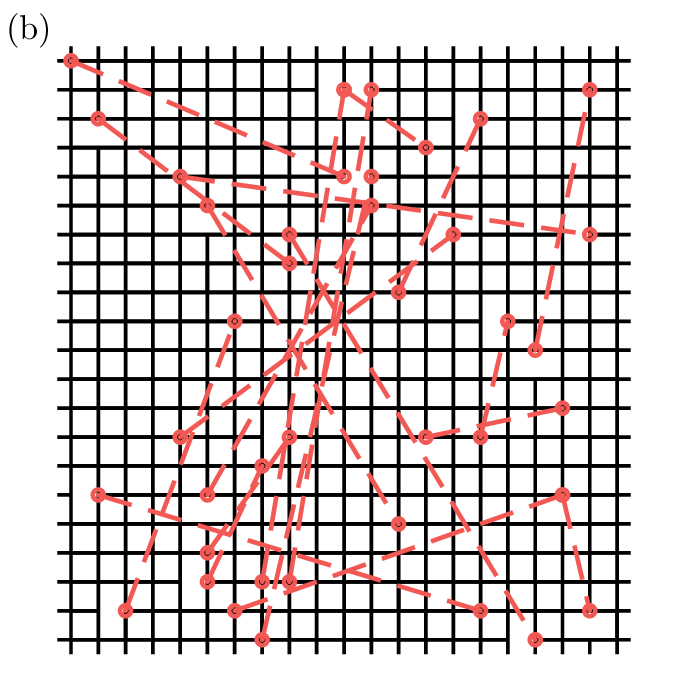}}
\end{subfigure}
\caption{ (Color online) (a) SSW network obtained from a square lattice with lattice spacing equal to unity and side $L=21$ by means of rewiring with probability $\phi=0.03$ within $R \in [1,5]$. The rewired bonds are shown as dashed red lines between two nodes shown by red (gray) circles. The bonds going over the boundary reappear on the other side due to periodic boundary conditions.
(b) SW network obtained from the square lattice by means of infinite-range bond rewiring with probability $\phi=0.03$.
}
\label{fig:rewiring}
\end{figure}

Let us consider a network consisting of $N$ nodes arranged on a regular two-dimensional lattice with each node connected to the same number of nearest neighbours, $q>2$.
In particular, we studied honeycomb ($q=3$), square ($q=4$) and triangular ($q=6$) lattices in which the bonds connecting nearest neighbours can be rewired with probability $\phi$ to a randomly chosen node under the constraint of no self- or double bonds.
The probability of rewiring was assumed to be independent of the states of the nodes (cf. Refs.~\cite{Gross_2006:PRL,Rattana_2014:PRE}).
Two types of models for bond rewiring were considered:
(i) Spatial small-world (SSW) networks with rewiring to a random node within a finite distance, $R \in [R_{\text{min}}, R_{\text{max}}]$, where $R_{\text{min}}>0$ and $R_{\text{max}}$ are parameters of the model which are assumed to be independent of the lattice size (see Fig.~\ref{fig:rewiring}(a));
(ii) SW networks with rewiring to any random node within the system (see Fig.~\ref{fig:rewiring}(b)) which is the limiting case of a spatial-SW if $R_{\text{min}}=1$ and $R_{\text{max}}(L) \sim L \to \infty$. Here, $L$ is the linear size of the system.
In SSW networks, the rewiring is local which contrasts with SW networks where it is global.
The bond rewiring was performed in the following way.
Consider for concreteness a square lattice (see Fig.~\ref{fig:rewiring}).
It can be constructed by an $N$ times repeated translation of a node with two bonds attached to it (horizontal and vertical), over a distance $a=1$ in both the horizontal and vertical direction producing $N=L\times L$ nodes on a square grid.
These two bonds attached to each node are then rewired with probability $\phi$ (per bond) to a random node within the range, $[R_{\text{min}},R_{\text{max}}]$, subject to no double bonds.
Such a bond-rewiring algorithm similar to that used in Ref.~\cite{Watts_Strogatz:1998} does not preserve the degree distribution, in contrast to degree-preserving algorithms used in analysis of homogeneous SW networks~\cite{Maslov2002,Santos2005}, and it results in  degree distribution~\cite{Barrat_EPJB2000}, which differs from that of a random graph (see App.~\ref{app:B} for more detail).

The dynamics of the SIR process is defined by the life-time $\tau=1$ of any node in the infected state and rate $\lambda_{ij}$ of stochastic transmission of infection from node $i$, infected at time $t_i$, to an attached node $j$ in the S-state.
The SIR process with simple non-synergistic Poisson transmission is described by a constant transmission rate, $\lambda_{ij}=\alpha$.
This model has been extensively studied on networks and it is well established that an SIR epidemic spreads (invades the network) if the transmission rate is greater than a critical value, $\alpha \ge \alpha_c$, which marks the epidemic or, equivalently, invasion threshold~\cite{Grassberger_83,Murray_02:book,Marro_99:book,Barrat_08:book}.
In the presence of synergy, the transmission rate of infection from an infected node, $i$, to a susceptible neighbour, $j$, is a piece-wise constant function. Step-wise changes occur after infection or removal events involving nodes $i$, $j$ or any other in their neighbourhood.  Such changes can be conveniently incorporated in numerical simulations using the event-driven continuous-time algorithm described in App.~\ref{app:AA}.

Our aim was to investigate the effects of synergistic transmission and rewiring (both varying probability and range of rewiring) on the invasion threshold.
We focused on a specific type of synergy associated with the number of infected neighbours of a susceptible recipient node (referred to as \emph{r}-synergy in~\cite{Perez_Reche_2011:PRL}).
In this case, the individual transmission of infection from node $i$ to node $j$
occurs with the rate $\lambda_{ij}(t)$ which depends on the number $n_j(t)$ of infected neighbours of recipient node $j$ excluding the attacker $i$~\cite{Perez_Reche_2011:PRL,Taraskin-PerezReche_PRE2013_Synergy}.
The individual transmission between node $i$ and $j$ starts at time $t_{i}$ when the node $i$ was infected and it stops at
$t_i'$, the time of infection of node $j$, not necessary by node $i$, or the time of recovery of node $i$, i.e. $t_i'=t_i+\tau$.
The number  $n_j(t)$ of infected neighbours of $j$ can vary with time in a step-wise manner for $t\in [t_i, t_i')$.
The time locations of the steps correspond to the stochastic infection and deterministic recovery events for the neighbours of node $j$ excluding $i$ and they depend on the history of the system
at $t < t_i$.

In particular, we analysed invasions for the following three functional forms of $\lambda_{ij}(t)$ given as implicit functions of $n_j(t)$:
\begin{itemize}
\item[(i)] Exponential rate,
\begin{eqnarray}
\lambda_{ij}(t)= \alpha e^{\beta n_j(t)} ~,
\label{eq:rate_exp}
\end{eqnarray}
\item[(ii)] the linear approximation to the exponential rate in Eq.~\eqref{eq:rate_exp},
\begin{equation}
\lambda_{ij}(t) = (\alpha + \alpha \beta n_j(t)) \theta(1 + \beta n_j(t))~,
\label{eq:rate_linear1}
\end{equation}
\item[(iii)] and a linear rate~\cite{Perez_Reche_2011:PRL,Taraskin-PerezReche_PRE2013_Synergy},
\begin{equation}
\lambda_{ij}(t) = (\alpha + \beta'n_j(t))\theta(\alpha + \beta'n_j(t))~.
\label{eq:rate_linear2}
\end{equation}
\end{itemize}
Here, the Heaviside function takes the value $\theta(x)=1$ for $x\ge 0$ and is zero, otherwise.
The expressions for these rates are valid for $t \in [t_i,t_i']$. For times outside this interval, $\lambda_{ij}=0$.

\begin{figure}[ht]
\centering
\begin{subfigure}[b]{0.3\textwidth}
\centering
{\includegraphics[clip=true,width=\textwidth]{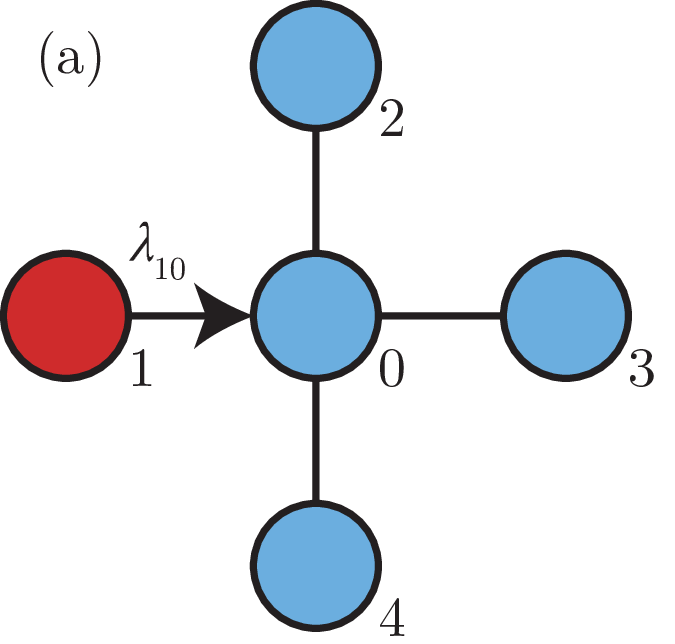}}
\end{subfigure}
\quad
\begin{subfigure}[b]{0.3\textwidth}
\centering
{\includegraphics[clip=true,width=\textwidth]{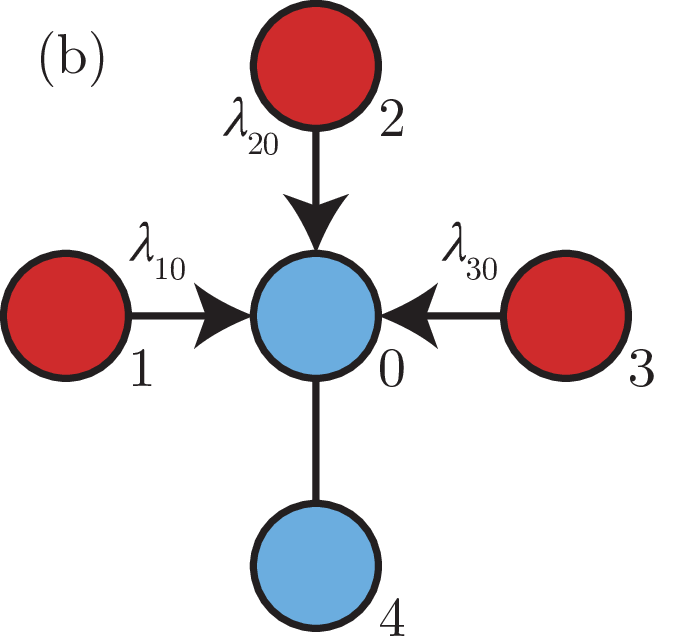}}
\end{subfigure}
\caption{ (Color online) Schematic illustration of (a) a non-synergistic attack from infected node 1 (red or gray) to susceptible node 0 (blue or light gray) when all other
nearest neighbours (nodes 2, 3, 4) of node 0 are susceptible (blue or light gray) and
(b) synergistic attacks from nodes 1, 2 and 3, all infected (red or gray), to susceptible node 0.
The synergy effect in the case of simultaneous attack from three nodes in (b) is taken into account by a change in transmission rate from $\lambda_{10}=\alpha $ in the case of a single non-synergistic attack in (a) to $\lambda_{10}=\lambda_{20}=\lambda_{30}=\alpha e^{2\beta}$ for the exponential form given by Eq.~\eqref{eq:rate_exp}.
}
\label{fig:synergy_rates}
\end{figure}

The rate $\alpha$ in Eqs.~\eqref{eq:rate_exp}-\eqref{eq:rate_linear2} refers to the inherent (synergy-free) transmission rate.
The coefficient $\beta$ in Eqs.~\eqref{eq:rate_exp} and \eqref{eq:rate_linear1} accounts for constructive ($\beta>0$) or destructive ($\beta < 0$) synergy and is assumed to be independent of $\alpha$.
If $\beta>0$ ($\beta<0$) then the rate $\lambda_{ij}(t)$ can exceed (be smaller than) the inherent rate $\alpha$ in the presence of a finite number of infected neighbours of node $j$ (see Fig.~\ref{fig:synergy_rates}).
The choice of the exponential function ensures positive values of the transmission rate for all values of $\beta$.
In the linear approximation to the exponential rate, the synergy contribution is proportional to the product of the inherent rate and synergy coefficient, i.e. $\propto \alpha \beta$.
However, it is possible that the synergistic effects do not depend on the inherent rate which is described by the functional form given by Eq.~\eqref{eq:rate_linear2} where the synergy rate, $\beta'$, is independent of $\alpha$.

\section{Numerical results}
\label{sec:Results}

In order to study the effects of the model parameters ($\beta$ or $\beta'$, $R_{\text{min}}$, $R_{\text{max}}$, and $\phi$) on the invasion threshold, $\alpha_{\text{c}}$,  we numerically analysed the SIR process with synergistic rates defined by Eqs.~\eqref{eq:rate_exp}-\eqref{eq:rate_linear2} on both SSW and SW networks with periodic boundary conditions.
All our simulations correspond to a linear size $L\alt 200$ for the underlying lattices.
The transmission dynamics were modelled as a continuous-time synchronous process by
using kinetic Monte-Carlo simulations  as described in App.~\ref{app:AA}.

 For any given set of parameters, the value of $\alpha_{\text{c}}$ was estimated by
 using finite-size scaling analysis for one-dimensional spanning epidemics~\cite{PerezReche2003,PerezReche_PRL2008,Perez_Reche_2011:PRL} in the case of SSW networks with finite-range rewiring, $R_{\text{max}}\ll L$. For SW networks with infinite-range rewiring, we used a linear fit~\cite{Sander_02} and/or inflexion point for the mass of the infinite cluster {\it vs} inherent transmission rate~\cite{Newman_99:PRE,Newman_2002:PRE} (see App.~\ref{app:A} for details).

\begin{figure}[ht]
\centering
\begin{subfigure}[b]{0.45\textwidth}
\centering
{\includegraphics[clip=true,width=\textwidth]{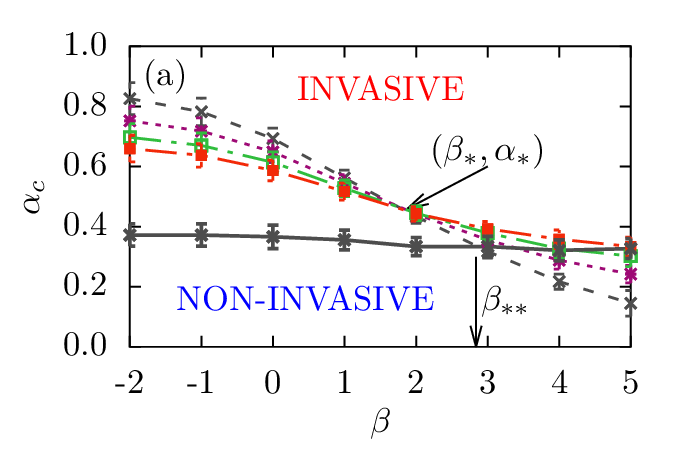}}
\end{subfigure}
\quad
\begin{subfigure}[b]{0.45\textwidth}
\centering
{\includegraphics[clip=true,width=\textwidth]{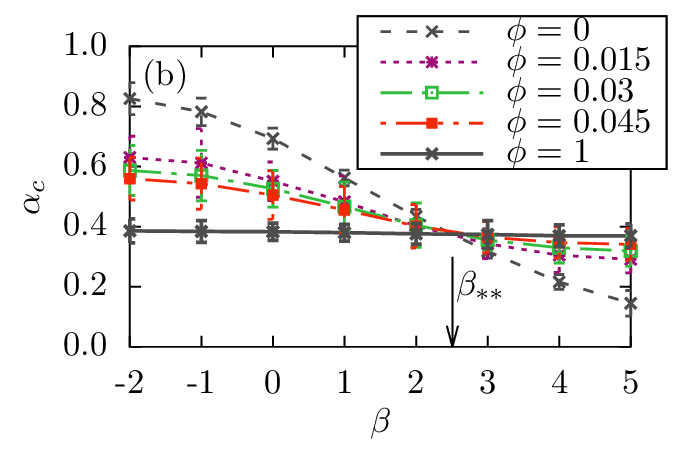}}
\end{subfigure}
\caption{ (Color online)  Invasion phase diagrams (inherent critical rate, $\alpha_c$, {\it vs} synergy parameter, $\beta$) for SIR epidemics on (a) a SSW built from a square lattice ($q=4$) with rewiring range $R \in [1,13]$, \\
(b) a SW network with infinite-range rewiring on a square lattice.
For both finite- and infinite-range rewiring, the exponential form of the transmission rate given by Eq.~\eqref{eq:rate_exp} was used.
Different line styles refer to the phase boundaries corresponding to different values of $\phi$ as marked in the figure legend.
The crossing points of the phase boundaries for the two limiting cases of complete (solid lines for $\phi=1$) and no (dashed lines for $\phi=0$) rewiring occur at
 $\beta=\beta_{**}$ where $\beta_{**}\simeq 2.9 \pm 0.3 $ in (a) and $\beta_{**} \simeq 2.5 \pm 0.5$ in (b).
}
\label{fig:alpha_vs_beta_phi}
\end{figure}
\begin{figure}[ht]
\centering
\begin{subfigure}[b]{0.45\textwidth}
\centering
{\includegraphics[clip=true,width=\textwidth]{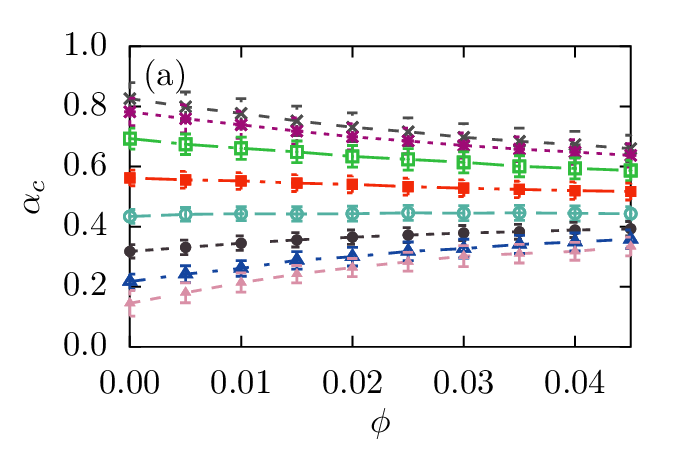}}
\end{subfigure}
\quad
\begin{subfigure}[b]{0.45\textwidth}
\centering
{\includegraphics[clip=true,width=\textwidth]{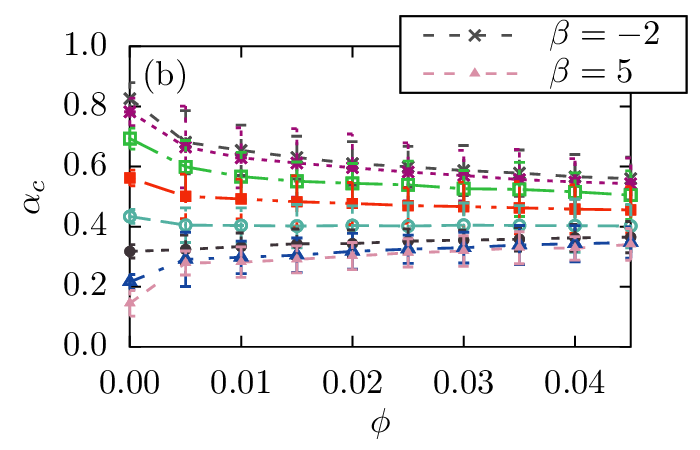}}
\end{subfigure}
\caption{ (Color online)
The dependence of the critical inherent transmission rate on rewiring probability $\phi$ for different values of $\beta$ between $\beta=5$ (lowest curve) and $\beta=-2$ (uppermost curve) with the other curves corresponding to values of $\beta$ incremented by unity (the lines are shown to guide the eye only).
Panels (a) and (b) show data for the same models as in Fig.~\ref{fig:alpha_vs_beta_phi}.
The horizontal lines correspond to $\alpha_{\text{c}} \simeq \alpha_*$ and $\beta \simeq \beta_*$ with $(\beta_*,\alpha_*)\simeq (1.88 \pm 0.04 , 0.46 \pm 0.02)$ in panel (a) and
$(\beta_*,\alpha_*) \simeq (2.03\pm 0.03, 0.40\pm 0.03)$ in panel (b).
}
\label{fig:alpha_vs_phi}
\end{figure}

Fig.~\ref{fig:alpha_vs_beta_phi} shows how the critical value of the inherent transmission rate depends on the synergy parameter $\beta$ for the exponential rate given by Eq.~\eqref{eq:rate_exp} on a square lattice ($q=4$). Results are shown for SSW (panel (a)) and SW (panel (b)) networks.
Each line in the figures gives the invasion threshold $\alpha_c$ as a function of $\beta$ for given $\phi$ and rewiring range.
SIR epidemics in systems with $\alpha$ and $\beta$ above/below the invasion line are invasive/non-invasive.
For fixed values of $\phi$, the critical transmission rate $\lambda_c$ depends on two parameters $\alpha_c$ and $\beta$ in such a way (see Eqs.~\eqref{eq:rate_exp}-\eqref{eq:rate_linear2}) that if $\beta$  increases,  then the value of $\alpha_c$ should decrease in order to keep the same value of $\lambda_c$.
Therefore, as expected, $\alpha_c$ decreases with increasing $\beta$ for any fixed $\phi<1$, meaning that increasing synergistic cooperation systematically makes systems less resilient to epidemic invasion.
The effect of rewiring probability on $\alpha_c$ is more involved and the trend depends on the value of $\beta$. For small rewiring probability ($\phi\ll 1$), all the phase-separation lines intersect
at a single model-dependent point $(\beta_*,\alpha_*)$.
This means that for a certain value of $\beta=\beta_*$, the critical inherent transmission rate $\alpha_c=\alpha_*$ does not depend on the rewiring probability $\phi$ (see horizontal lines in Fig.~\ref{fig:alpha_vs_phi} and dependence of the mass of the infinite cluster on $\alpha$ in Fig.~\ref{fig:mass_vs_alpha}(b) in App.~\ref{app:A}).
For values of $\beta<\beta_*$, the critical threshold in $\alpha$ decays with increasing $\phi$ (see Fig.~\ref{fig:alpha_vs_phi} and Fig.~\ref{fig:mass_vs_alpha}(a)).
This is the expected behaviour for synergy-free epidemics in SW networks~\cite{Moore_00}.
In contrast, for relatively strong synergy, $\beta>\beta_*$, the critical value of $\alpha$ increases with rewiring probability (see Fig.~\ref{fig:alpha_vs_phi} and Fig.~\ref{fig:mass_vs_alpha}(c)).
In other words, the more bonds rewired in the system, the more resilient it becomes to strongly synergistic SIR epidemics.
This finding is in qualitative agreement with the social reinforcement effects discussed by Centola~\cite{Centola_2010:Science}. It is important to note, however, that this effect does not only require the synergistic effects to be constructive (i.e. $\beta>0$) but it also requires that they are strong enough so that $\beta > \beta_*>0$.

\begin{figure}[ht]
\centering
\begin{subfigure}[b]{0.45\textwidth}
\centering
{\includegraphics[clip=true,width=\textwidth]{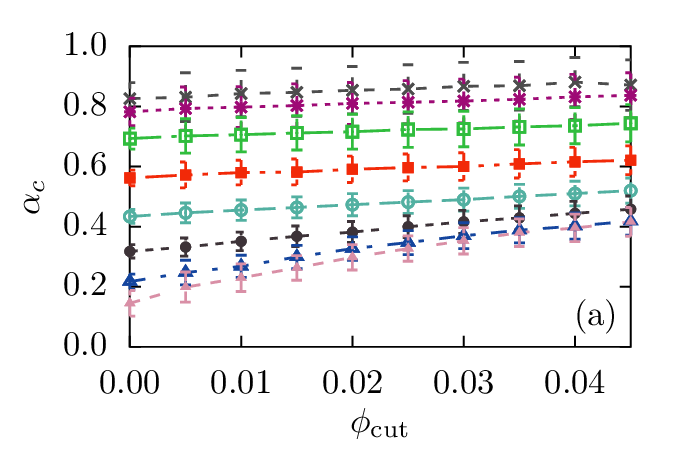}}
\end{subfigure}
\quad
\begin{subfigure}[b]{0.45\textwidth}
\centering
{\includegraphics[clip=true,width=\textwidth]{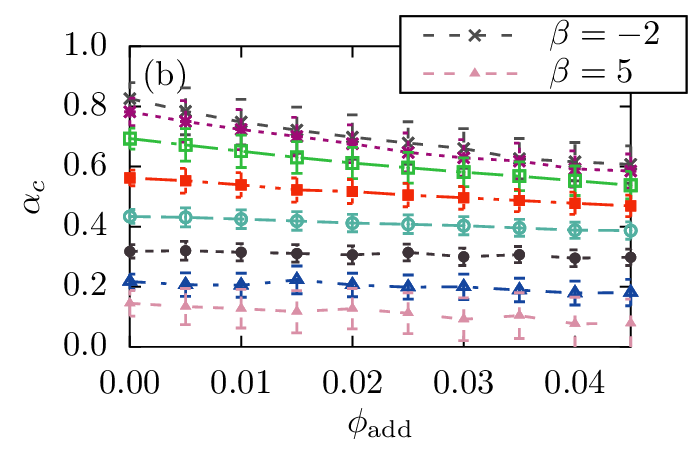}}
\end{subfigure}
\caption{ (Color online)
The dependence of the critical inherent transmission rate on the probabilities (a) $\phi_{\text{cut}}$ of just cutting local bonds and (b) $\phi_{\text{add}}$ of just adding global bonds for different values of $\beta$ on a square lattice (same line styles as in Fig.~\ref{fig:alpha_vs_phi}). The exponential form of the transmission rate given by Eq.~\eqref{eq:rate_exp} was used for obtaining the data presented in both panels and bonds were added in the finite range $R\in[1,13]$ for the model in (b).
}
\label{fig:add_remove}
\end{figure}

In order to interpret the results presented above, it is helpful to analyse separately the effects of cutting local bonds and adding local or global connections which are the two basic operations involved in rewiring.
For large positive $\beta$, removing
 short-range bonds alone increases the resilience of the system whilst adding bonds on its own~\cite{Newman_2002:PRE,Sander_02} decreases its resilience to SIR epidemics (see Fig.~\ref{fig:add_remove}).
In the rewiring scenario, these two tendencies compete and the resulting effects on invasion depend on the synergy strength.
The effect of enhanced resilience with increasing rewiring is observed only for relatively large values of the synergy parameter $\beta$ when the constructive synergy helps the SIR process to evolve locally where support from the infected neighbours is strongest.
The rewired bonds diminish the local connectivity and bring the infection to such remote parts of the system where there are practically no infected nodes which could support further spread of infection.
The loss in ability to spread locally is more significant than the gain due to jumps to remote places where the advantages of high constructive synergy cannot be used efficiently.
This is the reason why addition of new rewired shortcuts can make the system more resilient.

\begin{figure}[ht]
\centering
\quad
\begin{subfigure}[b]{0.45\textwidth}
\centering
{\includegraphics[clip=true,width=\textwidth]{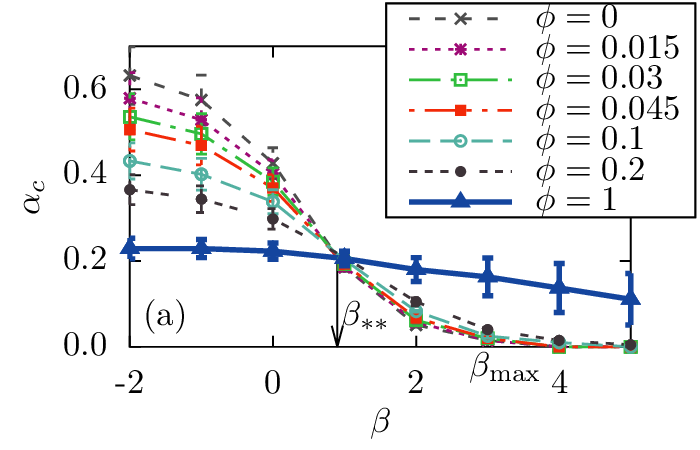}}
\end{subfigure}
\quad
\begin{subfigure}[b]{0.45\textwidth}
\centering
{\includegraphics[clip=true,width=\textwidth]{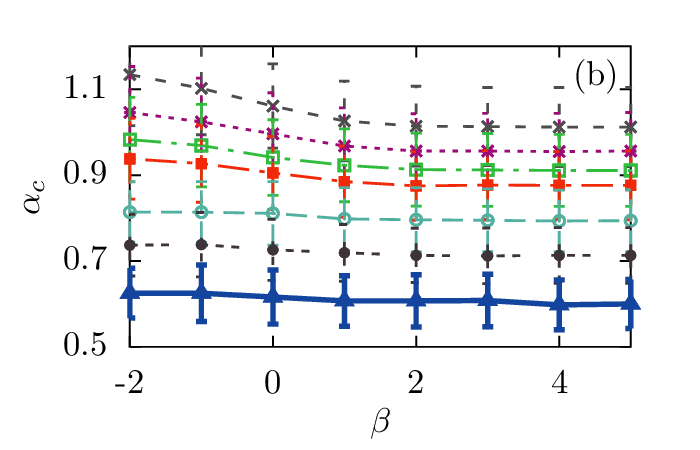}}
\end{subfigure}
\caption{ (Color online) Invasion phase diagrams for (a) triangular ($q=6$) and (b) honeycomb ($q=3$) lattices with finite-range rewiring with $R \in [1,13]$. An exponential rate given by Eq.~\eqref{eq:rate_exp} was used. The same line styles as in panel (a) for $\phi \in [0,1]$ were used in panel (b). The crossing point in (a) between the phase boundaries for networks without rewiring ($\phi=0$) and complete rewiring ($\phi=1$) corresponds to $\beta=\beta_{**}\simeq 0.9 \pm 0.1$.
}
\label{fig:triangular_hexagonal}
\end{figure}

The arguments presented above relied on the interesting property of the intersection point $(\beta_*,\alpha_*)$ which does not significantly depend on rewiring with $\phi \ll 1$. However, the conclusions remain valid for any value of $\phi$. In general, the intersection point of the phase boundaries for arbitrary $\phi \in (0,1)$ and for $\phi=0$ occurs at a point $(\hat{\beta}(\phi),\hat{\alpha}(\phi))$ which depends on $\phi$. The function $\hat{\alpha}(\phi)$ monotonically decreases with $\phi$ from $\hat{\alpha}(0)=\alpha_{*}$ to $\hat{\alpha}(1)=\alpha_{**} < \alpha_*$. In contrast, $\hat{\beta}(\phi)$ is a monotonically increasing function taking values between $\hat{\beta}(0)=\beta_{*}$ and $\hat{\beta}(1)=\beta_{**}>\beta_{*}$. These trends can be seen in Figs.~\ref{fig:alpha_vs_beta_phi}(a) and \ref{fig:triangular_hexagonal}(a) for networks with underlying square and triangular lattices, respectively.

Considering a fully rewired lattice with $\phi=1$ allows the effect of enhanced resilience to SIR epidemics with strong synergy to be predicted independently from the analysis of SW networks.
Indeed, for $\phi=1$, the network becomes similar to a random (Erd\"os-R\'enyi) graph where local lattice connections are rare, resulting in an absence of small loops.
Therefore, in this limit, the simultaneous presence of more than one infected node in the neighbourhood of the recipient is very unlikely and thus the synergy effects should be negligible.
In other words, the epidemic threshold does not depend significantly on $\beta$.
Consequently, the phase boundary is close to a horizontal line in the $\beta$-$\alpha$ plane and, importantly, this line can intersect the phase boundary for $\phi=0$ at the point $(\beta_{**},\alpha_{**})$ (see Fig.~\ref{fig:alpha_vs_beta_phi} and Fig.~\ref{fig:triangular_hexagonal}(a)).
This means that for $\beta\agt \beta_{**}$ the fully-rewired system is more resilient to invasion than the original lattice without rewiring. Again, this result contrasts with the prediction from models with simple transmission suggesting that, given a mean node degree, invasions are more likely in random graphs than in regular lattices~\cite{Watts_Strogatz:1998,Newman_99:PRE,Moore_00,Sander_02,Santos2005,Khaleque_2013,Grassberger_2013:J_Stat_Phys}.
For instance, the critical transmissibility, $T_c=1-e^{-\alpha_c}$ (a complementary quantity marking epidemic threshold, see Sec.~\ref{sec:Analytics} for more detail),    on a square lattice ($q=4$)  $T_c=1/2$ whereas  it is $T_c=1/(q-1)=1/3$ in a random graph with $\langle k\rangle =q=4$.  
Our model reproduces this traditional behaviour for $\beta < \beta_{**}$.
Such an effect becomes more pronounced for regular lattices with higher coordination number.
For example, in a triangular lattice ($q=6$), all the epidemics with exponential transmission rate are invasive for relatively large values of $\beta \agt \beta_{\text{max}} \simeq 4$.
However, in the case of a fully-rewired triangular lattice, the presence of an almost horizontal phase boundary at $\alpha_c = \tilde{\alpha} \simeq \alpha_{**}$ (with $\alpha_{**} \simeq 0.21 \pm 0.03$ for both the SW network and SSW network with $R \in [1, 13]$) clearly demonstrates that all the epidemics become non-invasive for $\alpha \alt \alpha_{**}$ including those which were invasive in non-rewired networks for $\beta \agt \beta_{\text{max}}$ (see Fig.~\ref{fig:triangular_hexagonal}(a)).

In lattices with relatively small coordination number (i.e with weak local connectivity), the effect of the rewiring-enhanced resilience to synergistic epidemics is not observed.
This is due to the fact that the effects of synergy in such lattices are not very strong and the phase boundaries in the lattices without rewiring are almost horizontal.
This effect is illustrated in Fig.~\ref{fig:triangular_hexagonal}(b) for the honeycomb lattice ($q=3$), where the invasion threshold is in the range, $1.02 \alt \alpha_c \alt 1.15$.
In the fully rewired lattice, the critical inherent rate takes values $\alpha_c =\tilde{\alpha} \simeq 0.57 \pm 0.07$ for infinite-range rewiring and $\alpha_c =\tilde{\alpha} \simeq 0.62 \pm 0.06$ for finite range rewiring with $R \in [1, 13]$.
In both cases, $\alpha_c$ is practically independent of $\beta$ and the invasion boundary is an almost horizontal line located at $\tilde{\alpha}$, below the range of the invasion boundary corresponding to $\phi=0$.
Therefore, it is not surprising that rewiring decreases the resilience of the system irrespective of the value of $\beta$ (see Fig.~\ref{fig:triangular_hexagonal}(b)).
These results show that the effect of local social reinforcement in networks with weak local connectivity
 is not strong enough to compete with the gain in the spread efficiency achieved by the shortcuts and rewiring in the networks. This effect is not captured by existing models with social reinforcement~\cite{Centola_2010:Science,Centola_2007:PhysicaA,Montanari_2010:PNAS,Lu_2011:NewJPhys,ZhengLuMing_2013_SocialReinforcement:PRE}.

The critical value of $\tilde{\alpha}$, defined by the position of the horizontal phase boundary in the $(\beta,\alpha)$ plane for fully rewired lattices, can be found in terms of the bond-percolation threshold~\cite{Grassberger_83}, $T_c$, as
\begin{eqnarray}
\tilde{\alpha} = \ln(1-T_c)^{-1}
~,
\label{eq:alpha_tilde}
\end{eqnarray}
where
$T_c=\langle k\rangle/\langle k(k-1)\rangle$~\cite{Cohen_2000:PRL,Callaway_2000:PRL,Newman_02:epidemic}.
The values of $\tilde{\alpha}$ obtained numerically for infinite-range fully rewired lattices with triangular ($ \tilde{\alpha}\simeq 0.21 \pm 0.03$), square ($\tilde{\alpha} \simeq 0.35 \pm 0.04$), and honeycomb ($ \tilde{\alpha}\simeq 0.57 \pm 0.07$) geometries agree well with the values calculated from Eq.~\eqref{eq:alpha_tilde} of $\tilde{\alpha} \simeq 0.201 \pm 0.001$, $\tilde{\alpha}\simeq 0.336\pm 0.002$ and $\tilde{\alpha} \simeq 0.558\pm 0.005$ for the same lattice types, respectively.

\begin{figure}[ht]
\centering
\includegraphics[clip=true,width=\textwidth]{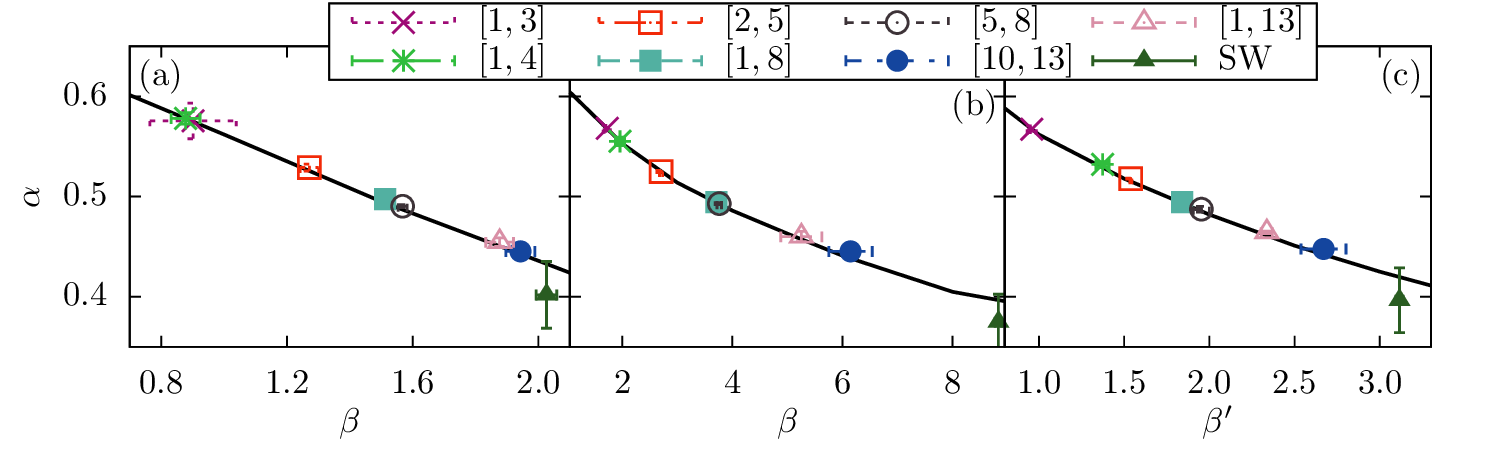}
\caption{ (Color online).
The set of characteristic parameters $(\beta_*,\alpha_*)$ for several SSW models and the SW model (with small probability of rewiring, $ \phi \le 4.5\times 10^{-2}\ll 1$) defined on a square lattice with (a) exponential (Eq.~\eqref{eq:rate_exp}), (b) linear approximation to the exponential (Eq.~\eqref{eq:rate_linear1}) and (c) linear (Eq.~\eqref{eq:rate_linear2}) transmission rates of infection for SIR epidemics.
In each panel, the solid line shows the invasion threshold separating non-invasive (below the line) and invasive (above the line) regimes in networks without rewiring ($\phi=0$). In panel (a), the continuous line corresponds to part of the curve in Fig.~\ref{fig:alpha_vs_beta_phi}.
Different symbols refer to different models of rewiring within the range $R\in [R_{\text{min}},R_{\text{max}}]$ as marked in the figure legend.
 }
\label{fig:all_models}
\end{figure}

The effect
 of rewiring-enhanced resilience to invasion does not change qualitatively for the variety of the models given sufficiently high coordination number of underlying lattice.
In particular, we found it for different ranges of rewiring including relatively small ones.
Similarly, the effect was observed for models with linear transmission rates given by Eqs.~\eqref{eq:rate_linear1} and \eqref{eq:rate_linear2}.
Fig.~\ref{fig:all_models} summarises our findings for the variety of models investigated.
Here, we show the location of three sets of characteristic points $(\beta_*,\alpha_*)$ found for different models with exponential (Eq.~\eqref{eq:rate_exp}), linear approximation to the exponential (Eq.~\eqref{eq:rate_linear1}) and linear (Eq.~\eqref{eq:rate_linear2}) transmission rates defined on square lattice.
These points, as expected (see Sec.~\ref{sec:Analytics}), belong to the corresponding phase separation lines, $\alpha_c(\beta,\phi=0)$, for models  without rewiring ($\phi=0$).

It should be noticed that the position of the characteristic point $(\beta_*,\alpha_*)$ changes in a systematic way, moving down along the phase boundary (for lattices without rewiring, see solid curve in Fig.~\ref{fig:all_models}) with increasing rewiring range, tending to the point corresponding to infinite rewiring range.
This can be understood as follows.
For given $\beta$ and $\phi \ll 1$, increasing the maximum rewiring radius, $R_{\text{max}}$, will increase the global connectivity whilst the local connectivity remains similar.
This has the effect of making the system more susceptible to invasion and thus, the larger the rewiring range, the smaller the critical inherent rate.
Therefore, the crossing point with the monotonically decaying curve corresponding to $\alpha_c(\beta,\phi=0)$ shifts downward along this curve.

\section{Analytical results for a model without correlations in transmission}
 \label{sec:Analytics}

Synergistic epidemics with removal can be viewed as correlated dynamical bond-percolation~\cite{Perez_Reche_2011:PRL}. In this mapping, the bond probability between two nodes $i$ and $j$ corresponds to the probability that node $i$ infects $j$ during its infectious period, $\tau$,
\begin{equation}
T_{ij}=1-\exp\left(-\int_{t_i}^{t_i+\tau} \lambda_{ij}(t) \text{d} t\right)~,
\end{equation}
where $t_i$ is the infection time of node $i$. The probability $T_{ij}$ is usually called the transmissibility from node $i$ to node $j$. Synergy makes the transmission rate $\lambda_{ij}(t)$ dependent on the infection history of the pair $i-j$ and its neighbourhood. Accordingly, transmissibility will be in general different for different pairs of hosts, i.e. transmissibility is heterogeneous over the set of pairs of hosts. This heterogeneity is annealed, i.e. it varies  with time, as opposed to quenched heterogeneity which is well studied for  epidemics~\cite{kuulasmaa1982,Cox_88,Sander_03,Kenah_07,Miller_07,Miller_JApplProbab2008,Neri_2011:JRSInterface,Handford_JRSInterface2011}, and its effect on the spread of epidemics is not obvious. In addition, transmissibilities of sufficiently close donor-recipient pairs have common nodes in their neighbourhoods and are not independent from each other if transmission is synergistic (since the infection history of close donor-recipient pairs neighbourhoods can overlap).
In Ref.~\cite{Perez_Reche_2011:PRL}, we showed that correlations in transmissibility can have a significant effect on invasion for large synergy. In spite of that, we found that the main features of invasion phase diagrams on lattices can be qualitatively described by a model which ignores spatial correlations but accounts for crucial spatial heterogeneity in transmissibilities~\cite{Perez_Reche_2011:PRL,Taraskin-PerezReche_PRE2013_Synergy}.
Here, we extend this approach to obtain approximate analytical results that explain the rewiring-enhanced resilience reported above.

 The critical transmissibility in rewired networks, $T_{\text{c}}(\phi)$, (i.e. the bond-percolation threshold) coincides with the mean transmissibility, $\langle T(\phi) \rangle$, in the system:
\begin{eqnarray}
\langle T(\phi) \rangle = T_{\text{c}}(\phi)~ .
\label{eq:condition}
\end{eqnarray}
The value of $T_c(\phi)$ depends on the topology of the network through the rewiring probability, but it does not depend on $\alpha$ or $\beta$ ($\beta'$).
The expression for the mean transmissibility, $\langle T(\phi) \rangle$, involves
averaging over degree distribution,
\begin{eqnarray}
\langle T(\phi) \rangle = \sum_k p_k \langle T_k \rangle ~,
\label{eq:T_mean_q}
\end{eqnarray}
and averaging over possible challenge histories of recipients with fixed number $k$ of nearest neighbours, accounted for by $\langle T_k \rangle$ in Eq.~\eqref{eq:T_mean_q}.

Eq.~\eqref{eq:condition} is valid for heterogeneous transmissibilities~\cite{Cox_88,Sander_02,Newman_02:epidemic} but it assumes the absence of correlations in transmissibilities for different bonds, which is true for non-synergistic SIR processes with a fixed removal time~\cite{Kenah_07}.
As argued above, such correlations are inherent for spread of the synergistic SIR process and condition~\eqref{eq:condition} does not hold in general~\cite{kuulasmaa1982,Kenah_07,Miller_JApplProbab2008,Neri_2011:JRSInterface}.
However, assuming that Eq.~\eqref{eq:condition} holds even for synergistic SIR processes leads to a quantitatively correct invasion phase diagram for small values of $\beta$ ($\beta'$) and a qualitatively correct picture for relatively large values of $\beta\sim 1$ ($\beta'\sim \alpha$)~\cite{Perez_Reche_2011:PRL,Taraskin-PerezReche_PRE2013_Synergy}.
In order to analytically study the consequences of Eq.~\eqref{eq:condition}, we linearise the dependence of $\langle T(\phi) \rangle$ and $T_{\text{c}}(\phi)$ on $\phi$. This leads to
an approximate condition for epidemic threshold which
 reads as (see App.~\ref{app:B} for more detail),
\begin{eqnarray}
T_{\text{c}0}(q) - A_q \phi &=&
1 -e^{-\alpha}\left(1-s_q(\alpha,\beta',0)B(\beta')\right)^{q-1}
\nonumber \\
&-&
 2\phi e^{-\alpha}\left[
(1-s_{q-1}(\alpha,\beta',0)B(\beta'))^{q-2} + (1-s_{q+1}(\alpha,\beta',0)B(\beta'))^{q}
\right]
\nonumber \\
&+&
 4\phi e^{-\alpha}\left(1-s_q(\alpha,\beta',0)B(\beta')
\right)^{q-1}
\nonumber \\
&+&
 (q-1)\phi e^{-\alpha}B(\beta')\frac{\partial s_q(\alpha,\beta',0)}{\partial \phi} \left(1-s_q(\alpha,\beta',0)B(\beta')
\right)^{q-2}
~,
\label{eq:critical_r2}
\end{eqnarray}
where $T_{\text{c}0}(q)$ is the bond-percolation threshold for a regular lattice with coordination number $q$ and the non-negative functions $s_q(\alpha,\beta',\phi)$, $B(\beta')$ and $A_q$ are introduced in App.~\ref{app:B}.

Eq.~\eqref{eq:critical_r2} can be solved for $\phi(\alpha,\beta')$ resulting in,
\begin{eqnarray}
\phi= \frac{T_{c0}(q)-\left[
1 -e^{-\alpha}\left(1-s_q(\alpha,\beta',0)B(\beta')\right)^{q-1}\right]}{F_q(\alpha,\beta')}~.
\label{eq:critical_r3}
\end{eqnarray}
if $F_q(\alpha,\beta')\ne 0$, where
\begin{eqnarray}
F_q(\alpha,\beta')= A_q-e^{-\alpha}&&\Bigg[
2(1-s_{q-1}(\alpha,\beta',0)B(\beta'))^{q-2}
+2(1-s_{q+1}(\alpha,\beta',0)B(\beta'))^q
-4(1-s_q(\alpha,\beta',0)B(\beta'))^{q-1}
\nonumber
\\
&+& (q-1) B(\beta')(1-s_q(\alpha,\beta',0)B(\beta'))^{q-2}\frac{\partial s_q(\alpha,\beta',0)}{\partial \phi}
\Bigg]
~.
\label{eq:F}
\end{eqnarray}
For a fixed value of $\beta'$, Eq.~\eqref{eq:critical_r3} defines how the critical value of the inherent transmission rate varies with rewiring probability.

For synergy-free epidemics with $\beta'=0$, the values of $s_q=0$ and $F_q(\alpha,\beta'=0)=A_q$. Accordingly, the critical inherent rate decreases with rewiring probability,
\begin{eqnarray}
\alpha_c=\alpha_{c0} - \ln\left(1+\frac{A_q \phi}{1-T_{c0}} \right)
\simeq \alpha_{c0} - \frac{A_q \phi}{1-T_{c0}}
~,
\label{eq:alpha_c}
\end{eqnarray}
where $\alpha_{c0}= -\ln(1-T_{c0})$ is the critical transmission rate in the lattice without rewiring and synergy.
However, for increasing synergy, the decay of $\alpha_c$ with increasing rewiring probability becomes less pronounced and eventually, depending on $s_k(\alpha,\beta')$, it can become an increasing function.
This happens at a characteristic value of $\beta'=\beta'_* $, when $\alpha_{\text{c}}=\alpha_*$ does not depend on $\phi$.
Within the linear approximation for $\phi \ll 1$, this
 is possible when
 both the numerator and denominator in Eq.\eqref{eq:critical_r3} are simultaneously equal to zero, i.e.
\begin{subequations}
\label{eq:alpha_beta_star}
\begin{eqnarray}
T_{c0}-\left[
1 -e^{-\alpha_*}\left(1-s_q(\alpha_*,\beta'_*,0)B(\beta'_*)\right)^{q-1}\right]&=0~,
\label{eq:alpha_beta_star_a}
 \\
F_q(\alpha_*,\beta'_*)&=0
\label{eq:alpha_beta_star_b}
~.
\end{eqnarray}
\end{subequations}
On the one hand, Eq.~\eqref{eq:alpha_beta_star_a}
gives the phase boundary for invasion in the absence of rewiring.
On the other hand, the condition imposed by Eq.~\eqref{eq:alpha_beta_star_b} ensures that the solution $(\beta_*,\alpha_*)$ does not depend on $\phi$, as was found numerically (see Fig.~\ref{fig:all_models}).

\begin{figure}[ht]
\centering
\begin{subfigure}[b]{0.45\textwidth}
\centering
{\includegraphics[clip=true,width=\textwidth]{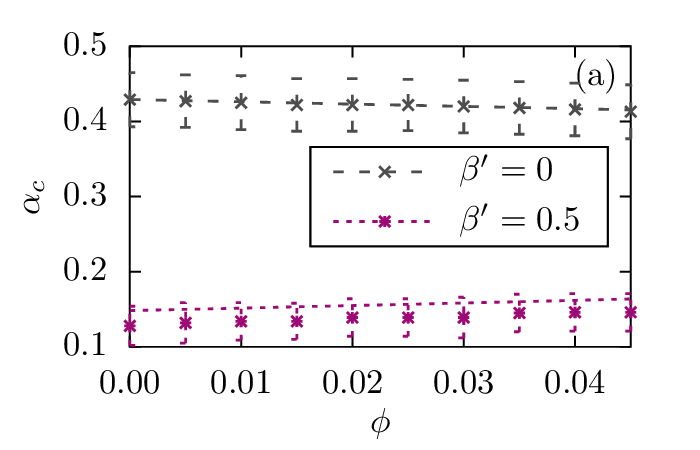}}
\end{subfigure}
\quad
\begin{subfigure}[b]{0.45\textwidth}
\centering
{\includegraphics[clip=true,width=\textwidth]{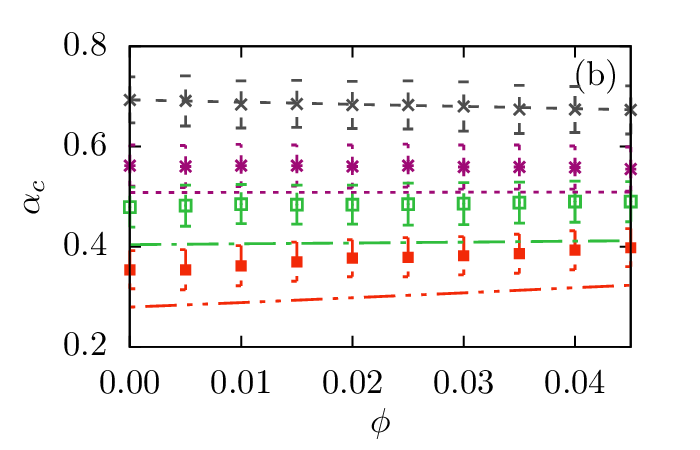}}
\end{subfigure}
\quad
\begin{subfigure}[b]{0.45\textwidth}
\centering
{\includegraphics[clip=true,width=\textwidth]{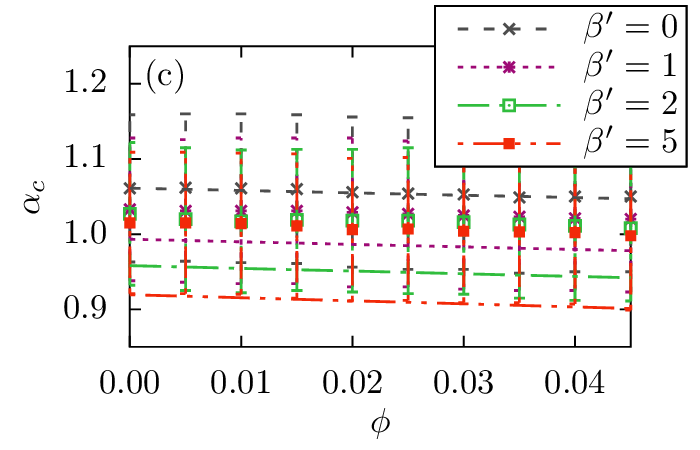}}
\end{subfigure}
\caption{ (Color online)  The dependence of the critical inherent transmission rate on the rewiring probability $\phi$ for (a) triangular ($q=6$), (b) square ($q=4$) and (c) honeycomb ($q=3$) lattices. The results were obtained for a linear transmission rate given by Eq.~\eqref{eq:rate_linear2} and rewiring within a finite range of $R \in [1,4]$. The symbols refer to the simulation results whilst the lines represent the model results given by Eq.~\eqref{eq:critical_r3}. The same symbols and line styles are used in panels (b) and (c).
The data for only two values of $\beta'$ are shown in (a) because $\alpha_c \to 0$ for $\beta'\agt 0.65$ (see Ref.~\cite{Taraskin-PerezReche_PRE2013_Synergy}).
}
\label{fig:alpha_vs_phi_model}
\end{figure}

For given $\beta'$, the solution of Eq.~\eqref{eq:critical_r3} for $\alpha_c(\phi)$ qualitatively agrees with numerical simulations. In the case of triangular (see Fig.~\ref{fig:alpha_vs_phi_model}(a)) and square (see Fig.~\ref{fig:alpha_vs_phi_model}(b)) lattices, the change from a decrease of $\alpha_c$ with increasing $\phi$ for $\beta' < \beta'_*$ to an increase for $\beta' > \beta'_*$ is observed.
At a characteristic value $\beta'=\beta'_*$, the inherent rate does not depend on $\phi$.
For honeycomb lattices (see Fig.~\ref{fig:alpha_vs_phi_model}(c)), no such transition is seen and the critical inherent rate only decreases with rewiring probability $\phi$.

The model is accurate for $\beta=0$ when there are no synergy effects and thus, no correlations in the transmission between different pairs of nodes (see top lines in Fig.~\ref{fig:alpha_vs_phi_model}).
Significant deviations between numerical data and model predictions are seen for larger values of $\beta \agt 1$ and they are due to approximations ignoring correlations in transmission. In spite of that, the model still provides qualitatively correct tendencies in $\alpha_c(\phi)$ for different values of $\beta'$ in various lattices.

Interestingly, in the case of just cutting or adding bonds, the condition given by Eq.~\eqref{eq:alpha_beta_star_b} is not satisfied for any $\beta'$, meaning that $\alpha_c$ varies with $\phi$ as a monotonically decreasing or increasing function for just cutting or adding bonds, respectively (see App.~\ref{app:C}).

\section{\label{sec:Conclusions} Conclusions}

To conclude, we have investigated the effects of local and global connectivity on spread of synergistic epidemics.
The underlying networks used in the analysis were two-dimensional lattices with different coordination number (honeycomb, square and triangular).
The local and global connectivity in these networks were changed by means of local (finite-range) and global (infinite-range) random bond rewiring.
The global bond rewiring produced two-dimensional small-world networks while the local rewiring created spatial small-world networks with geographical constraints on the finite length of rewired bonds.
 SIR epidemics with constructive and destructive synergy transmissions were analysed on such networks.
Our main findings are the following:
\begin{itemize}
\item[(i)] Bond rewiring enhances resilience to synergistic epidemics if two conditions are satisfied. First, the synergy effects are sufficiently strong and, second, the local connectivity is high enough. More specifically, the effect of rewiring-enhanced resilience is found only on lattices with high coordination number ($q\ge 4$) and synergy strength $\beta > \beta_*>0$. This finding is in line with those in Refs.~\cite{Centola_2010:Science,Centola_2007:PhysicaA,Montanari_2010:PNAS,Lu_2011:NewJPhys,ZhengLuMing_2013_SocialReinforcement:PRE}.
\item[(ii)] Independent of local connectivity, if constructive synergy is not strong enough, i.e. $ 0<\beta <\beta_*$, or synergy is destructive ($\beta <0$), rewiring enhances the spread of (reduces the resilience to) epidemics.
In other words, destructive and weakly constructive synergy do not change qualitatively behaviour of synergy-free epidemics in rewired (small-world) networks~\cite{Newman_99:PRE,Moore_00,Sander_02,Khaleque_2013,Grassberger_2013:J_Stat_Phys}. In particular, the fact that the traditional framework is recovered for $ 0<\beta <\beta_*$ challenges the statement of Refs.~\cite{Centola_2010:Science,Centola_2007:PhysicaA,Montanari_2010:PNAS,Lu_2011:NewJPhys,ZhengLuMing_2013_SocialReinforcement:PRE} showing that rewiring-enhanced resilience of epidemic invasion does not occur for every constructive synergistic mechanism.
\item[(iii)] Independent of the strength of the synergy (constructive or destructive), if the local connectivity is small enough, the rewiring always decreases the resilience of the network to SIR epidemics. In particular, we have demonstrated this effect for epidemics in rewired honeycomb lattices ($q=3$).
\end{itemize}

All these three effects are quite robust to changes in the functional form of synergy transmission rate.
In our approach, synergy is modelled by the dependence of the transmission rate of infection between a donor-recipient pair on the number of infected neighbours of the recipient.
Three types of functional dependence of the transmission rate on the number of infected neighbours of donor-recipient pairs were investigated. Similar
 effects of local and global connectivity on spread of synergistic epidemics were found for all of them.
The strength of synergistic effects was controlled with a single parameter, $\beta$, which allowed both constructive ($\beta>0$) and
destructive ($\beta<0$) synergy effects to be studied.
This might be considered as an advantage of our model relative to other approaches typically studying one type of synergy, either constructive or destructive~\cite{Centola_2007:PhysicaA,Montanari_2010:PNAS,Lu_2011:NewJPhys}.
The effects reported here correspond to small-world networks obtained with a rewiring strategy which brings heterogeneity in the node degree. However, such heterogeneity is expected to play a secondary role on synergistic effects (e.g. rewiring-enhanced resilience) compared to rewiring-induced changes in local and global connectivity. In particular, we expect similar interplay between synergy and local/global topology when using a rewiring strategy leading to small-world networks with homogeneous node degree \cite{Maslov2002,Santos2005}.

\begin{acknowledgments}
We would like to thank Will Jennings for participating in developing the code used in simulations.  
\end{acknowledgments}

\appendix
\section{\label{app:AA}Algorithm}
In this Appendix, we describe the rules of the SIR process and algorithm used in the simulations.

The SIR process can be described as a trajectory in discrete state space with the state vector having $N$ components which can have three discrete values corresponding to different states of the nodes (S, I, and R) resulting in  $3^N$ states in total, $\{{\cal S}_i\}$.
The process evolves by means of instantaneous jumps between the states ${\cal S}_i$.
These jumps occur at times $t_i$ (elapsed from the start of the process at $t=0$) when the system rests in state ${\cal S}_i$ and  the trajectory is an
ordered in time  sequence of states ${\cal S}_i(t_i)$.
The jumps can occur only between the states described by state vectors
which differ in  one  component only, i.e. only one node changes its individual state after the jump.
Only one of two changes are possible in one event: infection,  i.e. S$\to$I or removal, i.e. I$\to$R.

The time intervals between jumps are defined by the dynamical rules of the process, namely by the rules for infection and removal.
The removal rule states that a node becomes deterministically removed   after time $\tau$ (parameter of the model) elapsed since the moment of infection of this node.
For example, if node $j$ has been infected at time $t_i$ with the system being in state ${\cal S}_i$, it is removed at time $t_k=t_i+\tau$ when the system is in the state  ${\cal S}_k$.
The states ${\cal S}_i$ and ${\cal S}_k$ can be separated on the trajectory of the process by many other system  states  which correspond to infection and removal  of other nodes.
The individual transmission of infection from an infected node to a susceptible one connected to the infected node  occurs stochastically at times given by a Poisson process. The rate of such processes remains constant during quiescent intervals of time between consecutive states but, in the presence of synergy, its value can change after transition events.
For example, assume that the system jumps from the state ${\cal S}_{i-1}$ to the next state on the trajectory
${\cal S}_{i}$ at time $t_i$ and infection event occurs at time $t_i$, i.e. a susceptible node $k$ becomes infected.
The process of infection of node $k$ is a superposition of independent individual infection transmissions from all infected neighbours connected to $k$ and it takes place with the rate $\lambda_k = \sum_m \lambda_{mk}$ where $m$ runs over all infected neighbours of $k$.
The value of $\lambda_k$ does not depend on time for $t\in[t_{i-1},t_i)$ and it is fully defined by the state of the system ${\cal S}_{i-1}$ at $t=t_{i-1}$, i.e. it does not depend on the previous history of the system at $t< t_{i-1}$.
In particular, the rate $\lambda_k$ depends on the number of infected neighbours and on individual rates $\lambda_{mk}$.
In general, the rates $\lambda_{mk}$ can also depend
on the number of infected neighbours of node $k$
(only for  non-synergistic epidemics  the values of $\lambda_{mk}$ are  independent of the infected neighbours of $k$).
After the infection event at time $t=t_i$, infection rates between any infected-susceptible pairs of connected nodes may have changed and should be updated. Similarly,
if a deterministic removal rather than infection event  takes place at $t_i$ then all the individual infection rates should also be updated.

Numerically, we aimed to sample without bias all possible trajectories of the SIR process.
This can be achieved by means of kinetic Monte-Carlo~\cite{LandauBOOK} simulations exploiting the Gillespie algorithm (direct method~\cite{Gillespie_1976:J_Comp_Phys,Li_2008:BiotechnolProg}) with modifications accounting for deterministic recovery events.
Within this algorithm the SIR trajectory was sampled as follows.
\begin{itemize}
\item[1.] Start simulations at $t=0$ by infecting a small number of nodes, $N_0 \sim \text{O}(1) \ll N$, distributed randomly within the network.
Create a list of infection events, i.e. the list of susceptible nodes linked to the infected nodes and  cumulative infection rate (sum of all individual infection rates) for each node in the list.
Create a list of nodes in the infected state with their recovery times.
\item[2.] For a current time step $t$, calculate the cumulative infection rate, $R=\sum_{i,j} \lambda_{ij}$  where $i$ runs over all infected nodes and $j$ runs over susceptible neighbours of infected nodes connected by the links to them, i.e. the sum is evaluated over all possible individual infection transmissions in the network.
\item[3.] If $R>0$, calculate a  uniformly distributed random number $r_1 \in (0,1]$.
\item[4.] Calculate a time step till the next possible infection event, $\Delta t=-\ln(r_1)/R$.
\item[5.] Compare $t+\Delta t$ with the time of the earliest  deterministic recovery event,  $t_{\text{r}} (>t)$.
If $t+\Delta t \ge t_{\text{r}}$ or $R=0$ perform the recovery event at $t_{\text{r}}$, update the list of the individual infection rates, set the current time $t$ to $t=t_{\text{r}}$ and return to step 2. 
\item[6.] If $t+\Delta t < t_{\text{r}}$, calculate a uniform random number $r_2 \in (0,1]$.
\item[7.] Add the individual infection rates for nodes from the list of infection events cumulatively until it exceeds $r_2 R$. Infect the node for this event, update the infection rates and recovery times and set $t=t+\Delta t$.
\item[8.] Return to step 2.
\end{itemize}
The simulation stops when there are no nodes in infected state.
This algorithm is valid for both synergistic and synergy-free SIR processes.
For a synergy-free process, the infection rates $\lambda_{ij}=\alpha$  are identical for all the individual infection processes.
In case of synergy,  the individual infection rates $\lambda_{ij}$ entering the expression for $R$ depend on the neighbourhood of susceptible nodes as described in Sec.~\ref{sec:Model}. 

\section{\label{app:A}Scaling analysis}

In this Appendix, we present the results for the finite-size scaling analysis performed to find the invasion threshold in the case of finite- and infinite-range rewiring.

\begin{figure}[ht]
\centering
\begin{subfigure}[b]{0.45\textwidth}
\centering
{\includegraphics[clip=true,width=\textwidth]{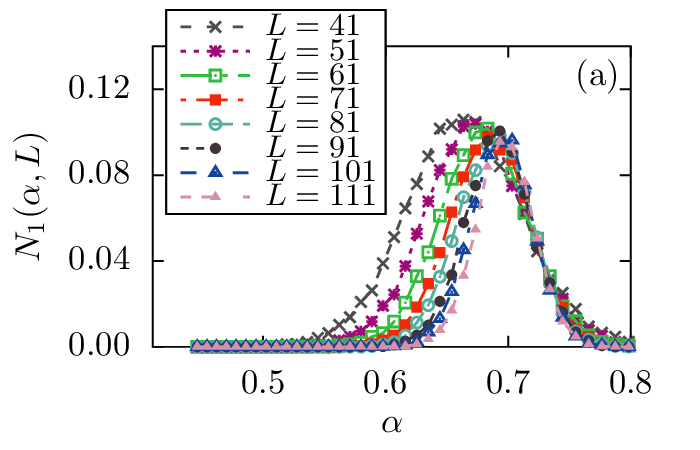}}
\end{subfigure}
\quad
\begin{subfigure}[b]{0.45\textwidth}
\centering
{\includegraphics[clip=true,width=\textwidth]{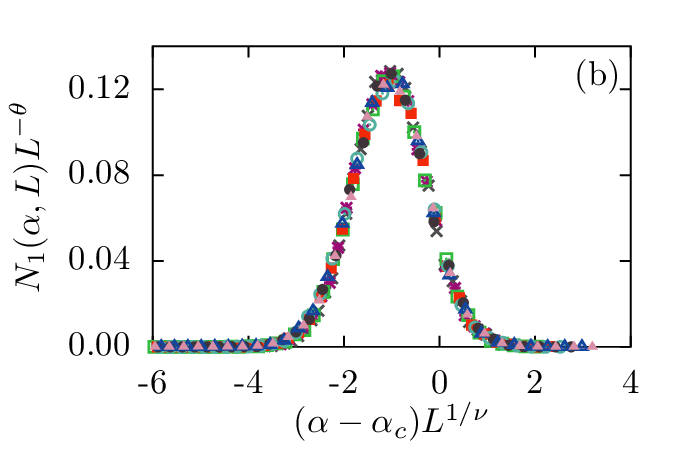}}
\end{subfigure}
\caption{ (Color online) (a) The relative number of one-dimensional spanning clusters, $N_1(\alpha, L)$, {\it vs} the inherent transmission rate, $\alpha$, for several lattice sizes, $L$, as indicated in the figure legend. (b) The scaling collapse of $N_1(\alpha,L)L^{-\theta}$ when plotted against $(\alpha-\alpha_{\text{c}})L^{1/\nu}$ with $\alpha_c=0.727$, $\nu \simeq 1.3\pm 0.1$ and $\theta=-0.0522$ for SIR epidemics on a square lattice with finite-range rewiring $R \in [1,8]$ with $\phi=0.025$ for an exponential form of the rate given by Eq.~\eqref{eq:rate_exp} with synergy parameter, $\beta=-1$. The different symbols in both panels correspond to different lattice sizes as indicated in the legend with each point averaged over 20000 realisations of the epidemics.
}
\label{fig:scaling}
\end{figure}

In order to find the invasion threshold, we exploit the fact that it corresponds to a critical point of the system.
Following this property, in models with finite-range rewiring, we determined the invasion threshold using  finite-size scaling for one-dimensional spanning clusters~\cite{PerezReche2003,PerezReche_PRL2008,Perez_Reche_2011:PRL,Taraskin-PerezReche_PRE2013_Synergy}.
The relative number of one-dimensional spanning clusters, $N_1(\alpha, L)$, exhibits a maximum near the critical value of the inherent transmission rate, $\alpha_c$ (see Fig.~\ref{fig:scaling}(a)), and the values of $N_1(\alpha,L)L^{-\theta}$ for varying $L$ should collapse onto a single master curve, $\tilde{N}_1(x)$, if plotted against $x=(\alpha-\alpha_{\text{c}})L^{1/\nu}$.
The exponents $\nu$ and $\theta$ and critical transmission rate, $\alpha_c$, are found from the scaling collapse (see Fig.~\ref{fig:scaling}(b)).
The value of the universal exponent $\nu \simeq 1.3 \pm 0.1$  is consistent with $\nu = 4/3$~\cite{Stauffer_92:book,Saberi_15:review} known for percolation in lattices without rewiring.
This is due to the restriction on the maximum rewiring distance in FSS to be much smaller and independent of the linear system size, $R_{\text{max}} \ll L$.

\begin{figure}[ht]
\centering
\begin{subfigure}[b]{0.45\textwidth}
\centering
{\includegraphics[clip=true,width=\textwidth]{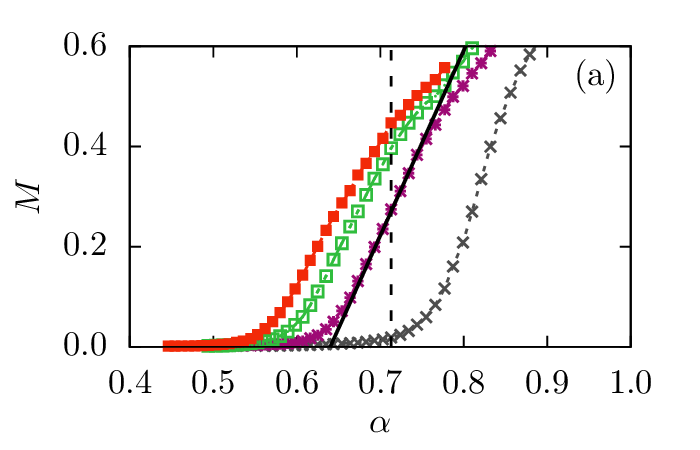}}
\end{subfigure}
\quad
\begin{subfigure}[b]{0.45\textwidth}
\centering
{\includegraphics[clip=true,width=\textwidth]{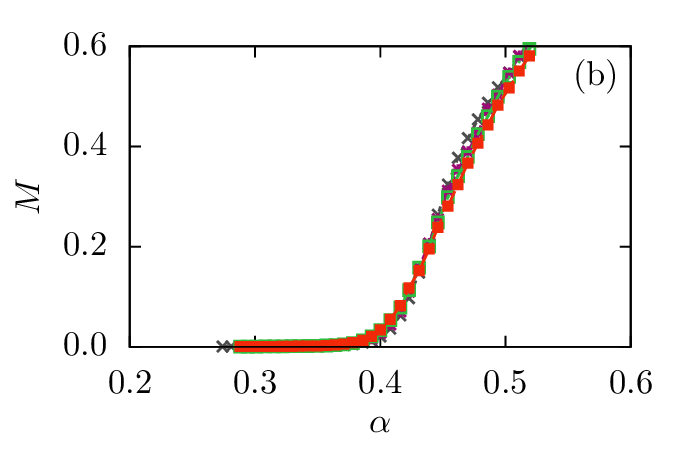}}
\end{subfigure}
\quad
\begin{subfigure}[b]{0.45\textwidth}
\centering
{\includegraphics[clip=true,width=\textwidth]{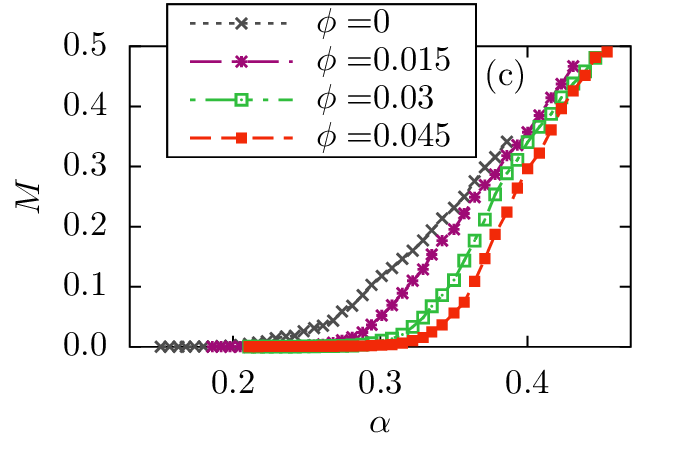}}
\end{subfigure}
\caption{ (Color online) The relative mass of the infinite cluster, $M$, for an SIR process on a square lattice of size $N=201 \times 201$ with infinite-range rewiring {\it vs} the inherent transition rate, $\alpha$, for different values of synergy parameter: (a) $\beta=-2$, (b) $\beta=2\simeq \beta_*$ and (c) $\beta=5$ and for varying rewiring probability, $\phi$, as indicated in the legends (the same symbols and line styles are used in all panels). An exponential form of the synergistic transmission rate given by Eq.~\eqref{eq:rate_exp} was used. An example of the linear fit used for estimating the critical inherent transmission rate is shown by the solid line in (a) for $\phi=0.015$ and the vertical dashed line goes through the inflection point for the $\phi=0.015$ curve.
}
\label{fig:mass_vs_alpha}
\end{figure}

The finite-size scaling used in the case of finite-range rewiring models cannot be applied to infinite-range rewiring.
This is due to the existence of an additional length scale related to the distance between the shortcuts~\cite{Ozana_2001:EPL,Newman_2002:PRE,Sander_02}.
Therefore, we used two complementary methods for estimating the critical threshold, $\alpha_c$.
The first method is based on the fact that small-world networks can be described by a mean-field approximation and thus, the relative mass of the infinite cluster, $M$, depends linearly on $(\alpha-\alpha_c)$ near the critical point.
A linear fit~\cite{Sander_02} was then used to estimate $\alpha_c$ (see solid line in Fig.~\ref{fig:mass_vs_alpha}(a)).
Alternatively, the critical value of inherent transmission rate can be found from the location of the inflection point on the curve for the mass of the infinite cluster, $M(\alpha)$~\cite{Newman_99:PRE,Newman_2002:PRE} (see vertical dashed line through the inflection point in Fig.~\ref{fig:mass_vs_alpha}(a)).
The big error bars seen in Figs.~\ref{fig:alpha_vs_beta_phi}-\ref{fig:alpha_vs_phi} are due to the limited world sizes available in the small-world simulations.
The position of the inflection point gives an upper estimate on the value of $\alpha_c$ whilst the linear fit provides a lower bound estimate.
The restricted system sizes are caused by limited processing power available which is required for large worlds with synergistic effects within the continuous-time Kinetic Monte-Carlo algorithm.
The different tendencies in $M(\alpha)$ with increasing rewiring probability, $\phi$, for small-world networks for different values of the synergy parameter are seen in Fig.~\ref{fig:mass_vs_alpha}: (a) $\beta<\beta_*$, the system becomes less resilient with increasing $\phi$; (b) $\beta \simeq \beta_*$, the mass of the infinite cluster practically does not depend on $\phi$; and (c) $\beta>\beta_*$, the system becomes more resilient with increasing $\phi$.

\section{\label{app:B}Technical details in analysis of analytical model}

In this appendix, first we derive an approximate condition for epidemic threshold given by Eq.~\eqref{eq:critical_r2} and then give a simplified version for it.

The condition for the epidemic threshold given by Eq.~\eqref{eq:condition} depends on the bond-percolation threshold and mean transmissibility.
To leading order in $\phi$ (with $\phi \ll 1$), the bond-percolation threshold for networks with rewiring in a finite range is given by,
\begin{eqnarray}
T_{\text{c}} (\phi)\simeq T_{\text{c}0}(q) - A_q(R_{\text{min}},R_{\text{max}}) \phi
~,
\label{eq:critical_T_vs_phi}
\end{eqnarray}
where $T_{\text{c}0}(q)$ is the bond-percolation threshold for a regular lattice and $A_q(R_{\text{min}},R_{\text{max}})>0$ is a model-dependent constant.

In order to obtain a linear approximation for the dependence of $\langle T(\phi) \rangle$ on $\phi$, we first derive the expression for degree distribution in a network with rewiring (following Ref.~\cite{Barrat_EPJB2000}) and then obtain an approximation for it in the  case of small $\phi \ll 1$.
For example, we consider the case of the square lattice ($q=4$) with bonds rewired according to the Watts-Strogatz  rewiring algorithm as described in Sec.~\ref{sec:Model}, i.e. a SW network. 
According to this algorithm, for each node, $Q=q/2$ out of $q=Q+Q_1$ bonds attached to this node are rewired with probability $\phi$.
This means that an arbitrary node has at least $Q$ bonds attached to it. 
The remaining $Q_1=q-Q$ bonds can be broken by rewiring, so that only $n_1$ out of $Q_1$ are still attached to the node. 
The random number $n_1$ is 
 distributed according to the binomial distribution, $B_{n_1}(Q,\phi)$.
In addition to $Q+n_1$ bonds, a random number $n_2$ of new bonds can be attached to the node as a result of bond rewiring from other nodes.
This number is also distributed with binomial distribution $B_{n_2}(NQ,\phi/N)$ with $NQ$ being the total number of bonds in the system (the terms $\sim \text{O}(N^{-1})$ were ignored).
Therefore, the node degree $k=Q+n_1+n_2$ distribution $p_k$ is given by,
\begin{eqnarray}
p_k(N,\phi) = \sum_{n=0}^{\text{min}(k-Q,Q)} B_n(Q,\phi) B_{k-Q-n}(NQ,\phi/N)~
,
\label{eq:p_k_exact}
\end{eqnarray}
if $k\ge Q$ and $p_k=0$ otherwise.
In the limit of large $N\gg 1$, the binomial distribution $B_{k-Q-n}(NQ,\phi/N)$ tends to the Poisson one and
\begin{eqnarray}
p_k(N,\phi) \to p_k(\phi) \simeq \sum_{n=0}^{\text{min}(k-Q,Q)} \binom{Q}{n} \phi^n (1-\phi)^{Q-n} \frac{(Q\phi)^{k-Q-n}}{(k-Q-n)!}e^{-Q\phi}~
,
\label{eq:p_k_approx}
\end{eqnarray}
which does not depend on the system size and coincides with the degree distribution obtained for a ring with nodes connected to $q$ nearest neighbours~\cite{Barrat_EPJB2000}.
The convergence of $p_k(N,\phi)$ to  $p_k(\phi)$, as demonstrated in Ref.~\cite{Barrat_EPJB2000}, is rather fast and for $N\sim 10^3$ the numerical data almost perfectly reproduce the limiting distribution given by Eq.~\eqref{eq:p_k_approx}.

The limiting case for small rewiring probabilities, $\phi \ll 1$, follows from Eq.~\eqref{eq:p_k_approx}.  In this limit, mainly nodes with coordination numbers $k=q-1, q$ and $q+1$ are present in the network and $p_k(\phi)$ is given by,
\begin{eqnarray}
p_k(\phi)\simeq (1-4\phi)\delta_{k,q}+2\phi\delta_{k,q-1}+2\phi\delta_{k,q+1}
~.
\label{eq:p_k}
\end{eqnarray}

The value of $ \langle T_k \rangle$ in the expression for mean transmissibility given by  Eq.~\eqref{eq:T_mean_q} takes into account the synergy effects, i.e. that the transmission of infection from a donor to a recipient can occur in the presence of different numbers of infected neighbours (excluding the donor) of the recipient ($n=1,2,\ldots, q-1$) which can affect the transmission rate and thus the transmissibility.
Within the model of a time-dependent environment with linear transmission rate given by Eq.~\eqref{eq:rate_linear2}, the mean transmissibility
with fixed node degree $k$ for $\beta' > -\alpha/(k-1)$ is given by~\cite{Taraskin-PerezReche_PRE2013_Synergy},
\begin{eqnarray}
\langle T_k \rangle
=1 - e^{-\alpha}\left(1-s_kB(\beta')\right)^{k-1}
~,
\label{eq:T_mean_total}
\end{eqnarray}
where $s_k=s_k(\alpha,\beta',\phi)$ (parameter of the model) is the probability that a neighbour of a recipient node (excluding the donor, i.e. one out of $k-1$ neighbours) has been infected within the time-interval $[-\tau,\tau]$ if the donor became infected and started to challenge the recipient at $t=0$.
The function
 $B(\beta')=1-(1-e^{-\beta'})/\beta'$ increases monotonically from $0$ to $1$ with increasing $\beta'\to\infty$.

Combining Eqs.~\eqref{eq:T_mean_q}, \eqref{eq:T_mean_total} and \eqref{eq:p_k} gives the desired linear approximation for $\langle T(\phi) \rangle$. Introducing this expression and Eq.~\eqref{eq:critical_T_vs_phi} into Eq.~\eqref{eq:condition} leads to an approximate condition for the invasion threshold, $\alpha=\alpha_{\text{c}}(\beta',\phi)$, given by Eq.~\eqref{eq:critical_r2} which can be transformed to  Eq.~\eqref{eq:alpha_beta_star}.

\begin{figure}[ht]
\centering
\begin{subfigure}[b]{0.45\textwidth}
\centering
{\includegraphics[clip=true,width=\textwidth]{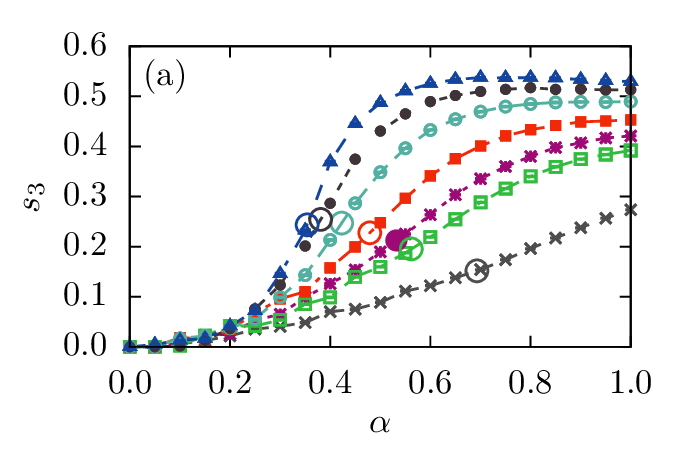}}
\end{subfigure}
\quad
\begin{subfigure}[b]{0.45\textwidth}
\centering
{\includegraphics[clip=true,width=\textwidth]{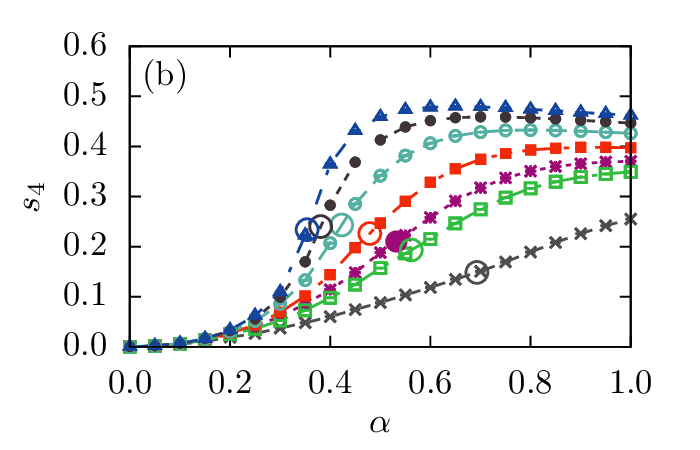}}
\end{subfigure}
\quad
\begin{subfigure}[b]{0.45\textwidth}
\centering
{\includegraphics[clip=true,width=\textwidth]{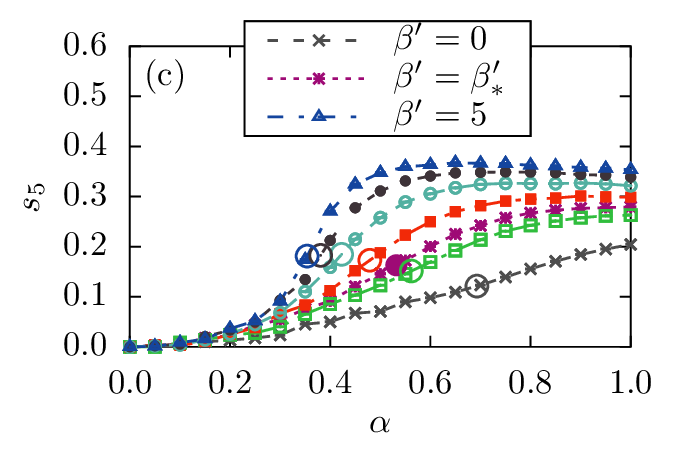}}
\end{subfigure}
\quad
\begin{subfigure}[b]{0.45\textwidth}
\centering
{\includegraphics[clip=true,width=\textwidth]{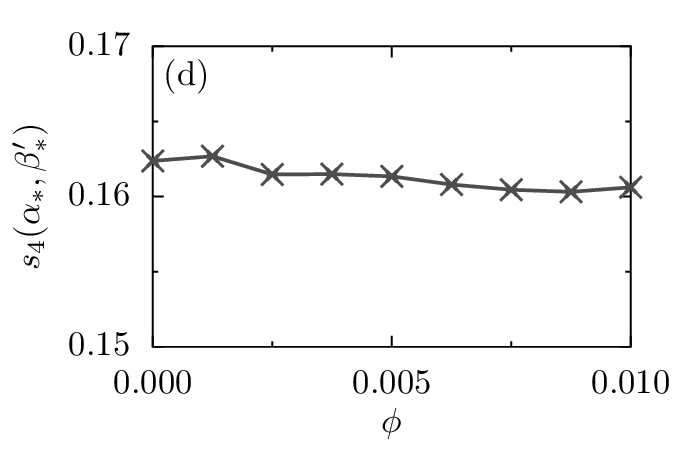}}
\end{subfigure}
\caption{ (Color online) Dependence of (a) $s_3$, (b) $s_4$ and (c) $s_5$ on inherent rate, $\alpha$, for different values of the synergy parameter, $\beta'$, (as marked in the legend in (c)) in square lattices with $\phi=0$ in (b) and a small value of rewiring probability $\phi=2.5\times 10^{-3}$ and $R \in [1,4]$ in (a) and (c). The unlabelled curves show values of $\beta'$ varying stepwise with unity from $\beta'=0$ (lowest curve) to $\beta'=5$ (topmost curve) except for the $\beta'_*$ curve. The figure in (d) gives the variation of $s_4(\alpha_*,\beta'_*)$ with rewiring probability $\phi$. The open circles in panels (a)-(c) represent the location of the critical inherent rate, $\alpha_c$, for each value of $\beta'$, whilst the solid circles show the location of $\alpha_*$.
}
\label{fig:s_vs_alpha}
\end{figure}

The solution of Eq.~\eqref{eq:alpha_beta_star_b} significantly depends on the functional form of the infection probabilities $s_q(\alpha,\beta',\phi=0)$ which can be found numerically in the same way as described in Ref.~\cite{Taraskin-PerezReche_PRE2013_Synergy}.
The dependence of $s_q(\alpha,\beta',\phi=0)$ on
$\alpha$ for $q=3$, $4$ and $5$ and different values of $\beta'$ are shown
in Fig.~\ref{fig:s_vs_alpha}.
It follows from this figure that $s_q(\alpha,\beta',\phi=0)\ll 1$ for all values of $q$ (at least for the values close to the invasion boundaries).

As a consequence, the function $s_q(\alpha,\beta',0)B(\beta') \ll 1$ and Eq.~\eqref{eq:alpha_beta_star_b} can be simplified as follows,
\begin{subequations}
\label{eq:alpha_beta_star_small_sB}
\begin{eqnarray}
T_{c0}-\left[
1 -e^{-\alpha_*}\left(1-(q-1)s_q(\alpha_*,\beta'_*,0)B(\beta'_*)\right)\right]&=0
\label{eq:alpha_beta_star_small_sB_a}
 \\
A_q-e^{-\alpha}B(\beta'_*)\Bigg[4(q-1)s_q(\alpha_*,\beta'_*,0)-2(q-2)s_{q-1}(\alpha_*,\beta'_*,0)-2qs_{q+1}(\alpha_*,\beta'_*,0)&+
\nonumber
\\
(q-1)\frac{\partial s_q(\alpha_*,\beta'_*,0)}{\partial \phi} \Bigg]&=0
\label{eq:alpha_beta_star_small_sB_b}
~.
\end{eqnarray}
\end{subequations}
Here, the derivative $ \partial s_q(\alpha_*,\beta_*,0)/\partial \phi $ is typically much smaller than $s_q(\alpha_*,\beta_*,0)$, as seen from  Fig.~\ref{fig:s_vs_alpha}(d).
If the expression in the square brackets is positive, Eq.~\eqref{eq:alpha_beta_star_small_sB_b} can be solved, with $(\alpha_*,\beta'_*)$ lying on the phase boundary for $\phi=0$ which follows from Eq.~\eqref{eq:alpha_beta_star_small_sB_a}.

\section{\label{app:C}Models with added and cut bonds}

In this Appendix, we study analytically and numerically two additional models with just adding and cutting the bonds instead of rewiring.

\begin{figure}[ht]
\centering
\begin{subfigure}[b]{0.45\textwidth}
\centering
{\includegraphics[clip=true,width=\textwidth]{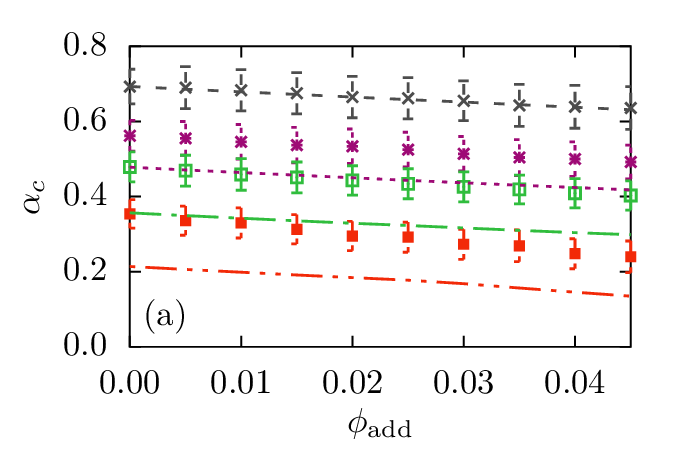}}
\end{subfigure}
\quad
\begin{subfigure}[b]{0.45\textwidth}
\centering
{\includegraphics[clip=true,width=\textwidth]{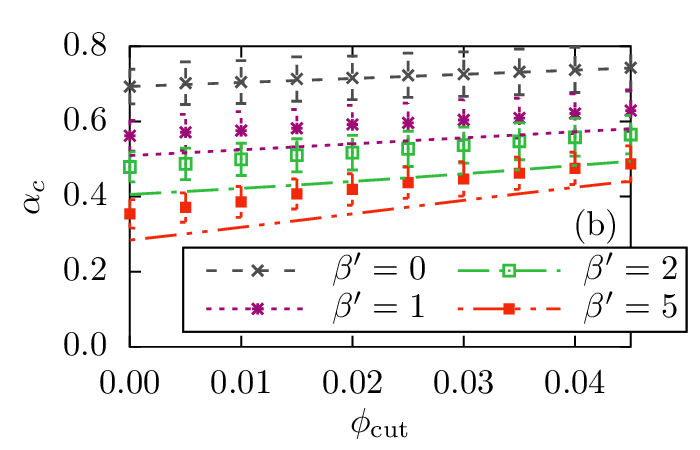}}
\end{subfigure}
\caption{ (Color online)  The dependence of the critical inherent transmission rate on the probabilities (a) $\phi_{\text{add}}$ of just adding global bonds  and (b) $\phi_{\text{cut}}$ of just cutting local bonds on a square lattice. The results were obtained for a transmission rate given by Eq.~\eqref{eq:rate_linear2} and adding the bonds within a finite range of $R \in [1,4]$. The symbols refer to the simulation results whilst the lines represent the model results given by Eq.~\eqref{eq:critical_r3} with $F_q(\alpha,\beta')$ for $q=4$ given by Eq.~\eqref{eq:F_add} in the left panel and Eq.~\eqref{eq:F_cut} in the right panel. The same values of $\beta'$ were used in both panels. 
}
\label{fig:alpha_vs_phi_add_cut}
\end{figure}

In a finite-range rewiring model, each bond can be rewired with probability $\phi$, i.e. the original bond is removed from the system and a new bond connecting two nodes within rewiring range is created.
Let us modify this model in such a way that the original bond is kept in the network in addition to the added bond.
This is a model with just added bonds~\cite{Newman_99:PRE,Moore_00}.
For small values of the probability of adding a bond, $\phi_{\text{add}} \ll 1$, the node degree distribution for this model is given by the following equation,
\begin{eqnarray}
p_q=(1-4\phi_{\text{add}})\delta_{q,4}+4\phi_{\text{add}}\delta_{q,5}~.
\label{eq:p_q_adding}
\end{eqnarray}
The corresponding expression for $F_q$ is given by
\begin{eqnarray}
F_q(\alpha,\beta')= A_q-e^{-\alpha}&&\Bigg[
4(1-s_{q+1}(\alpha,\beta',0)B(\beta'))^q
-4(1-s_q(\alpha,\beta',0)B(\beta'))^{q-1}
\nonumber
\\
&+& (q-1) B(\beta')(1-s_q(\alpha,\beta',0)B(\beta'))^{q-2}\frac{\partial s_q(\alpha,\beta',0)}{\partial \phi_{\text{add}}}
\Bigg]
\nonumber
\\
&\simeq&
A_q-e^{-\alpha}B(\beta')\left[4(q-1)s_q(\alpha,\beta',0)-4qs_{q+1}(\alpha,\beta',0)+(q-1)\frac{\partial s_q(\alpha,\beta',0)}{\partial \phi_{\text{add}}} \right]
~.
\label{eq:F_add}
\end{eqnarray}
The dependence of $\alpha_c(\phi)$ is given by Eq.~\eqref{eq:critical_r3} with
$F_q(\alpha,\beta')$ obeying Eq.~\eqref{eq:F_add}.
The results of its numerical solution are shown in Fig.~\ref{fig:alpha_vs_phi_add_cut}(a).
It can be seen from this figure that the critical inherent rate decreases with probability $\phi_{\text{add}}$ irrespective of the value of $\beta$. Such a monotonic trend is expected and agrees with the fact that Eq.~\eqref{fig:alpha_vs_phi_model}, which gives a necessary condition for $\alpha_c$ to be independent of $\beta$, is not satisfied for any $\beta$.
The analytical results are well supported by the results of the numerical simulations for $\phi_{\text{add}}\ll 1$ and $\beta \alt 1$.

Alternatively, the original rewiring model can be modified in such a way that
the original bonds are cut with probability $\phi_{\text{cut}}$ but new bonds are not added to the network.
This is a model with just removed bonds.
The node degree distribution for this model is given by the following equation,
\begin{eqnarray}
p_q=(1-4\phi_{\text{cut}})\delta_{q,4}+4\phi_{\text{cut}}\delta_{q,3}~.
\label{eq:p_q_cutting}
\end{eqnarray}
The corresponding expression for $F_q$ is given by
\begin{eqnarray}
F_q(\alpha,\beta')= A_q-e^{-\alpha}&&\Bigg[
4(1-s_{q-1}(\alpha,\beta',0)B(\beta'))^{q-2}
-4(1-s_q(\alpha,\beta',0)B(\beta'))^{q-1}
\nonumber
\\
&+& (q-1) B(\beta')(1-s_q(\alpha,\beta',0)B(\beta'))^{q-2}\frac{\partial s_q(\alpha,\beta',0)}{\partial \phi_{\text{cut}}}
\Bigg]
\nonumber
\\
&\simeq&
A_q-e^{-\alpha}B(\beta')\left[4(q-1)s_q(\alpha,\beta',0)-4(q-2)s_{q-1}(\alpha,\beta',0)+(q-1)\frac{\partial s_q(\alpha,\beta',0)}{\partial \phi_{\text{cut}}} \right]
~.
\label{eq:F_cut}
\end{eqnarray}
The dependence of $\alpha_c(\phi)$ is given by Eq.~\eqref{eq:critical_r3} with
$F_q(\alpha,\beta')$ obeying Eq.~\eqref{eq:F_cut}.
The results of its numerical solution are shown in Fig.~\ref{fig:alpha_vs_phi_add_cut}(b).
In this case, the critical inherent rate expectedly increases with $\phi_{\text{cut}}$.
The numeric results are again supportive of the analytics for relatively small values of $\beta$ and show qualitatively the same behaviour for $\beta\agt 1$.

\bibliographystyle{apsrev4-1.bst}

%


\end{document}